\renewcommand{\baselinestretch}{1}

\documentclass[9pt, nonacm, sigconf]{acmart}

\usepackage{graphicx}
\usepackage{booktabs}
\usepackage{multirow}
\usepackage{tabularx}
\usepackage{mathtools}
\usepackage{amsmath}
\usepackage{array}
\usepackage{algorithm}
\usepackage[noend]{algpseudocode}
\usepackage{xcolor}
\usepackage{enumitem}
\usepackage{threeparttable}
\usepackage{subcaption}
\usepackage{url}
\usepackage{float}
\usepackage{xcolor}
\usepackage{tikz,pgfplots}
\usetikzlibrary{patterns}
\usepackage[normalem]{ulem}

\usepackage[table]{xcolor}
\usepackage{balance}
\usepackage{nopageno}
\usepackage{verbatim}

\newcommand{\ignore}[1]{}
\newcommand{\redHL}[1]{\textcolor{red}{#1}}
\newcommand{\blueHL}[1]{{\textcolor{blue}{#1}}}

\definecolor{gray1}{gray}{0.90}

\newcommand\blfootnote[1]{%
  \begingroup
  \renewcommand\thefootnote{}\footnote{#1}%
  \addtocounter{footnote}{-1}%
  \endgroup
}

\usepackage{listings}
\newcommand{\eqLabel}[1]{(\refstepcounter{equation}\theequation)\label{#1}}

\begin{document}


\title{{COmPOSER:} \underline{C}ircuit \underline{O}ptimization of \underline{m}m-wave/RF circuits with \underline{P}erformance-\underline{O}riented \underline{S}ynthesis for \underline{E}fficient \underline{R}ealizations}

\author{
Subhadip Ghosh$^{1*}$, Surya Srikar Peri$^{1*}$, Ramprasath S.$^2$, Sosina A. Berhan$^1$, Endalk Y. Gebru$^1$, Ramesh Harjani$^1$, Sachin S. Sapatnekar$^1$ \\
$^1$\text{Department of Electrical and Computer Engineering, University of Minnesota, Minneapolis, MN, USA} \\
$^2$\text{Department of Electrical Engineering, Indian Institute of Technology Madras, Chennai, India}
}



\begin{abstract}
This work presents \textbf{COmPOSER}, an open-source, end-to-end framework for RF/mm-wave design automation that translates target specifications into optimized circuits with layouts. It unifies schematic synthesis, layout generation for actives and passives, and placement/routing, incorporating physics-based equations and machine-learning-driven electromagnetic models. 
Based on post-layout validation on multiple LNAs and PAs operating at up to 60~GHz in a commercial 65~nm process-kit, COmPOSER meets performance targets, comparable to expert manual designs, while delivering a 100--300$\times$ productivity gain.

\end{abstract}

\maketitle
\vspace{-2mm}
\section{Introduction}
\label{sec:Intro}
\noindent
Radio-frequency integrated circuits (RF ICs) form the backbone of modern wireless systems, from Wi-Fi and cellular networks to satellite and IoT platforms. RF and mm-wave circuits operate from a few GHz to over 30~GHz at wavelengths comparable to dimensions of on-chip interconnects and passives~\cite{sorin_book}.
{\blfootnote{\vspace*{-4pt}\noindent$^{*}$ These authors contributed equally to this work.}}
At high frequencies, parasitic coupling, substrate loss, routing geometry, and electromagnetic (EM) effects dominate circuit behavior, and circuit performance is tightly coupled to layout choices. As demonstrated in Section~\ref{sec:exp_results}, \textit{layout-oblivious optimization results in large errors: 25--33\% lower center frequency
and significant power and noise~figure degradation.}

The design of front-end blocks such as low-noise amplifiers (LNAs) and power amplifiers (PAs) must navigate these complexities as they simultaneously target conflicting objectives, e.g., gain ($G$), noise figure ($NF$), and bandwidth ($BW$), over a high-dimensional design space. 
Manual design has grown harder due to strong interdependencies between active and passive elements, alongside the nonlinear, nonunique layout-geometry to circuit performance mapping. This motivates the need for automated methodologies.

Traditional RF/mm-wave design has relied on manual and semi-structured approaches that codify established sizing, matching, and tuning practices into stepwise procedures~\cite{sorin_2007, taiyun_2023, lee_2002, chen_2012}, augmented with rules~\cite{mansour_2005}. While avoiding ad hoc design space exploration, they rely on designer intuition and do not incorporate layout parasitics. Early methods for automated schematic-level optimization were based on stochastic search~\cite{crols_95, gielen_2000, ricardo_2018}. Later formulations using Bayesian optimization~\cite{zeng_2018, zeng_2022}, convex programming~\cite{pileggi_2007}, and deep learning~\cite{ma_2022} continued to be layout-oblivious. 
Schematic-level methods do not account for layout parasitics that can dominate circuit behavior, and are simulation-intensive, with high computation costs.

To partially account for layout effects, several works have introduced EM-aware block-level optimization using ML surrogate models for inductors, transformers, and matching networks~\cite{zhang_2022, wang_2020, wang_2021, pan_2025}. These models improve accuracy only for passives, and require large datasets and long training cycles; data generation is a major bottleneck. In~\cite{sengupta_2023}, inverse design is used for mm-wave PA blocks, but focuses only on passives rather than the entire system.

Building on block-level methods, \textit{end-to-end flows} that couple sizing with layout generation, to reduce post-layout deviation, have been explored. At low frequencies, tighter integration has been demonstrated through circuit optimizers~\cite{Abel_19, Wang_2020_gcnrl, settaluri_20, ghosh_2025} and layout engines~\cite{kunal_2019, hao_2021, guo_2025}, but automation for RF/mm-wave circuits is limited. GASPAD~\cite{gielen_2014} for PAs and CYCLONE~\cite{sansen_2002} for VCOs add partial physical awareness using fixed templates or precharacterized parasitic abstractions but depend on expensive EM evaluations and are circuit-specific. Early schematic-layout co-design efforts~\cite{chi_2025, mendes_2025} do not include internal routing, pad routing, and power delivery network (PDN) routing. Given the strong impact of parasitics, their results are liable to fail performance constraints after these routing steps; hence, they are not truly end-to-end solutions.

To close the gap in available methods and the complexity of mm-wave systems, we introduce \textbf{COmPOSER}, an open-source framework for end-to-end specification-to-layout CMOS mm-wave/RF design automation. 
Its key advantages over prior approaches are:\\
\textbf{(1)}~COmPOSER provides a \textit{unified mm-wave/RF design framework} that achieves specification-to-GDSII automation, including device sizing, placement, routing, pad integration, and PDN synthesis.\\
\textbf{(2)}~COmPOSER combines circuit synthesis and physical realization within a \textit{single flow}.
It substitutes designer intuition with \textit{physics-informed equations} and incorporates \textit{precharacterized, technology-specific parasitics}. Since EM effects dominate transistor effects at high frequency, equation-based models incorporating parasitics are adequate for transistors; for passives, EM effects are captured accurately using \textit{ML-based inverse design}.\\
\textbf{(3)}~The COmPOSER methodology is \textit{transferable} across blocks: sizing is block-specific, but layout synthesis is block-agnostic.

\noindent
COmPOSER is applied to LNAs and PAs at up to 60~GHz. For a 40~GHz LNA, it shows \textbf{$>100 \times$} reduction in design turnaround time over manual design, with similar post-layout performance. 

\begin{figure}[b]
    \centering
    \setlength{\tabcolsep}{0pt} 
    \renewcommand{\arraystretch}{0} 
    \subfloat[]{%
        \includegraphics[width=0.37\linewidth]{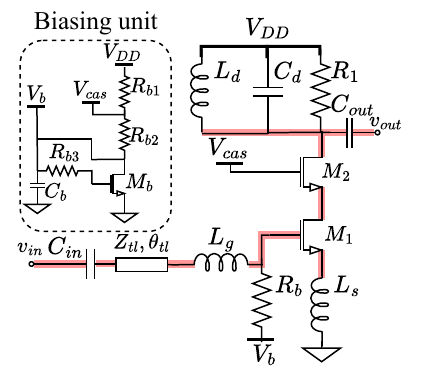}
         \vspace{-2pt}
        \label{fig:lna}
    }\hspace{2mm}
    \subfloat[]{%
        \includegraphics[width=0.47\linewidth]{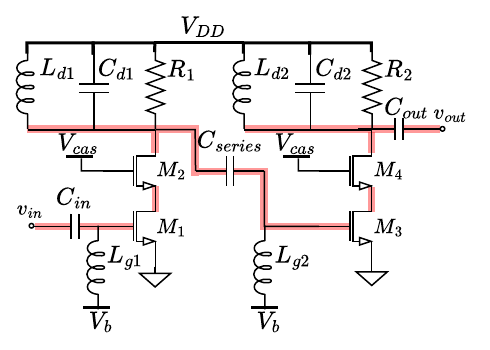}
         \vspace{-2pt}
        \label{fig:pa}
    }
    \vspace{-4pt}
    \caption{Topologies of
    (a) inductively-degenerated cascode LNA with bias network and 
    (b) two-stage Class-A PA.}
    \Description{Topologies of
    (a) inductively-degenerated cascode LNA with bias network and 
    (b) two-stage Class-A PA.}
    \label{fig:lna_pa}
\end{figure}

\begin{figure*}[t]
    \centering
    \includegraphics[width=0.85\textwidth]{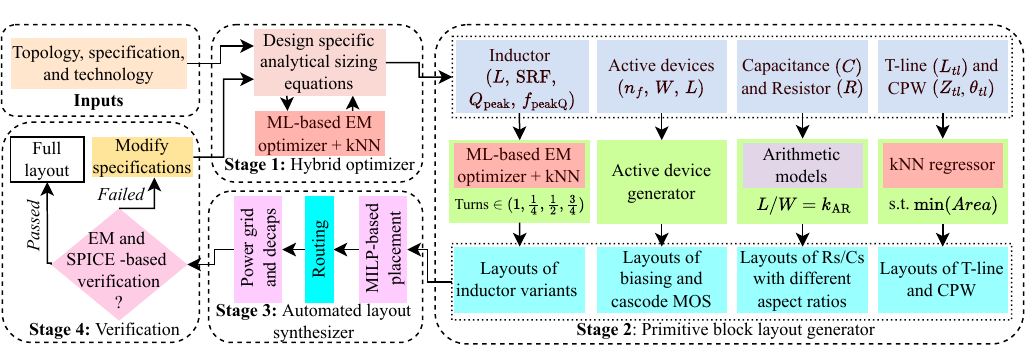}
    \caption{Overview of the proposed RF design automation framework: COmPOSER.}
    \Description{Overview of the proposed RF design automation framework: COmPOSER.}
    \label{fig:rf_flow}
    \vspace{-5mm}
\end{figure*}

\section{Manual Design vs. Proposed  Framework}

\label{sec:man_vs_composer}

\subsection{Challenges of manual design methodologies}
\label{sec:man_design}

\noindent
RF/mm-wave blocks such as LNAs, PAs, and passive matching networks are key components of transceivers. Their design requires the concurrent optimization of multiple blocks to meet performance targets under limited voltage headroom.
Conventional RF/mm-wave design follows a heuristic loop of schematic sizing, layout implementation, and EM validation, driven by designer experience. 

For the LNA and PA schematics in Fig.~\ref{fig:lna_pa}, the process begins with analytical biasing and sizing to approximately meet target specifications~\cite{razavi}. Each circuit employs input, output, and inter-stage (for multistage PA, Fig.~\ref{fig:lna_pa}(b)) matching networks, such as the gate, source, and drain inductors ($L_g$, $L_s$, $L_d$), capacitors ($C_{d_i}$, $C_{out}$), shielded transmission lines (T-lines) and coplanar waveguides (CPWs)~\cite{cpwref}. Together, they enable impedance transformation and conjugate matching for maximum power transfer and stable operation across the band. The design flow relies 
on a layout-oblivious step that sizes active and passive devices, followed by iterations that update parasitics through repeated layout, sizing updates, and electrical and EM simulation steps. These ad hoc iterations consume extensive design time and depend heavily on designer intuition.

Although LNAs and PAs serve different roles in the transceiver chain and have distinct performance metrics, they both rely on balancing device biasing, sizing, and load matching. The design process begins with device-level characterization, where unity-gain transit frequency ($f_T$), maximum oscillation frequency ($f_{max}$), and lowest achievable $NF$ ($NF_{min}$) are plotted as functions of drain current per micron of gate width to identify the optimal operating region. The optimal current density ($J_\mathrm{opt}$) is then chosen at the minimum of $NF_{min}$ for LNAs and the maximum of $f_{max}$ for PAs; these bias points are largely technology-dependent yet consistent across operating frequencies~\cite{sorin_book}. In an LNA, an inductively degenerated cascode provides low-noise input matching, with $L_s$ setting the $50~\Omega$ input impedance and $L_d$ resonating with device capacitance to maximize $G$. PAs follow the same bias–size–load principle but prioritize output power and efficiency, by employing Class-A or Class-AB stages 
refined through load-pull simulations to maximize power-added efficiency (PAE). Small layout variations and interconnect geometry alter parasitic loading and matching, necessitating repeated EM  simulations to recover target performance.

Among passive elements, on-chip inductors exert the strongest influence on RF and mm-wave performance. Their geometry dictates $G$, $NF$, and $BW$ through the peak quality factor ($Q_{peak}$), frequency of peak $Q$ ($f_{peakQ}$), and self-resonant frequency ($f_{SRF}$); all are highly layout-dependent and require full-wave EM simulations--which solve Maxwell’s equations on fine 3D meshes to capture coupling, substrate loss, and current crowding--for accurate evaluation. CPWs are characterized by their characteristic impedance ($Z_{cpw}$) and electrical phase length ($\theta_{cpw}$), introducing distributed effects that dominate beyond 10~GHz. Together, inductors and CPWs define the impedance environment and matching characteristics of LNAs and PAs, while capacitors complement them within matching networks. Resistors primarily support biasing and exhibit moderate layout sensitivity due to parasitic capacitance and substrate coupling at high frequencies. Decoupling capacitors (decaps) in the PDN further aid stability~\cite{niknejad_2008}, particularly under large-signal or fast-transient conditions, and also have modest layout dependence.

The strong coupling among devices, passives, and layout parasitics makes manual RF/mm-wave design slow, iterative, and dependent on designer intuition, and each schematic update demands fresh, expensive EM validation. 
These challenges motivate a unified, layout-aware automation framework that integrates analytical estimates with EM-informed models for specification-to-layout design.

\vspace{-0.5cm}
\subsection{Proposed \ignore{RF/mm-wave} design framework: COmPOSER}
\label{sec:proposed_framework}
\noindent
\textbf{COmPOSER} enhances the conventional RF/mm-wave design methodology by automating the key stages of the flow while retaining designer interpretability and control. As shown in Fig.~\ref{fig:rf_flow}, the framework converts target specifications into layout-ready designs through four tightly integrated stages, described in the rest of this section:  
\begin{itemize}[noitemsep, topsep=0pt, leftmargin=*]
    \item \textbf{Stage~1 (Hybrid optimizer)} combines analytical transistor and passive sizing. It employs an ML-based EM optimizer for EM-aware inverse design of inductors, coplanar waveguides, and transmission lines, accurate within 0.4\% of EM simulations, with $>1000\times$ speedup for this step. This EM optimizer is coupled with physics-based equations, enabling integrated EM-aware layout-level co-optimization of active and passive components.
    \item \textbf{Stage~2 (Primitive generator)} translates the optimized device and passive dimensions into compact, DRC-clean building blocks for active/passive components, generating geometric variants that facilitate efficient and constraint-aware placement and routing in later stages. 
    This enables hierarchical design, as well as scalability and adaptability across circuit blocks and topologies.
    \item \textbf{Stage~3 (Automated layout synthesizer)} integrates constraint-driven placement, formulated as a mixed-integer linear program (MILP)~\cite{wolsey_integer_programming_2020}, with A$^{*}$-based routing~\cite{hart_astar_1968} to construct low-parasitic interconnects, together with I/O placement and PDN generation for complete, implementation-ready layouts.
    \item \textbf{Stage~4 (Post-layout validation)} performs EM and circuit-level verification, followed by targeted resizing and layout tuning to recover any unmet performance metrics, ensuring convergence between schematic intent and physical implementation.  
\end{itemize}

\vspace{-2mm}
\section{Stage 1: Hybrid Optimizer}
\label{sec:stage_1}

\noindent
Stage 1 employs a hybrid optimizer, coupling physics-based MOS device equations with ML-based EM models for passives, enabling EM-aware co-optimization of transistor dimensions and passive geometries. The analytical model establishes device operating points and matching constraints, while EM models supply layout-accurate inductance and loss behavior. Due to the dominance of EM effects, analytical models are adequate for transistors, and we place high emphasis on fast, EM-accurate passive modeling and synthesis.

\noindent
\textbf{EM optimizer: random forest coupled with \ignore{k-nearest neighbor}k-NN.}
COmPOSER synthesizes inductors, T-lines, and CPWs \textit{in microseconds} using ML, avoiding costly full-wave EM runs used in manual design.  

\begin{figure}[h]
    \centering
    \setlength{\tabcolsep}{0pt}  
    \begin{tabular}{@{}cccc@{}}
        \includegraphics[width=0.25\linewidth]{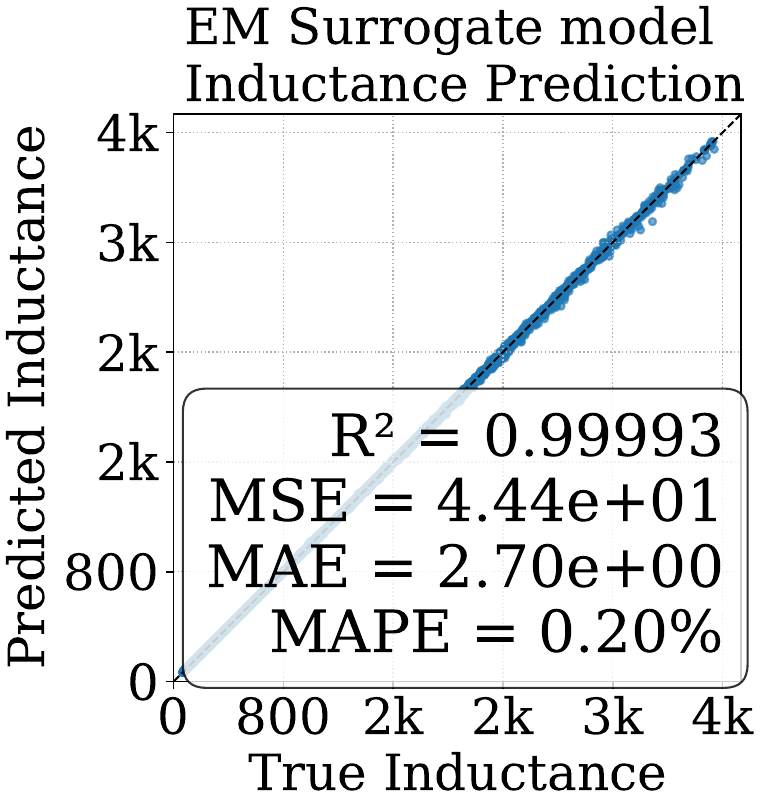} & 
        \includegraphics[width=0.24\linewidth]{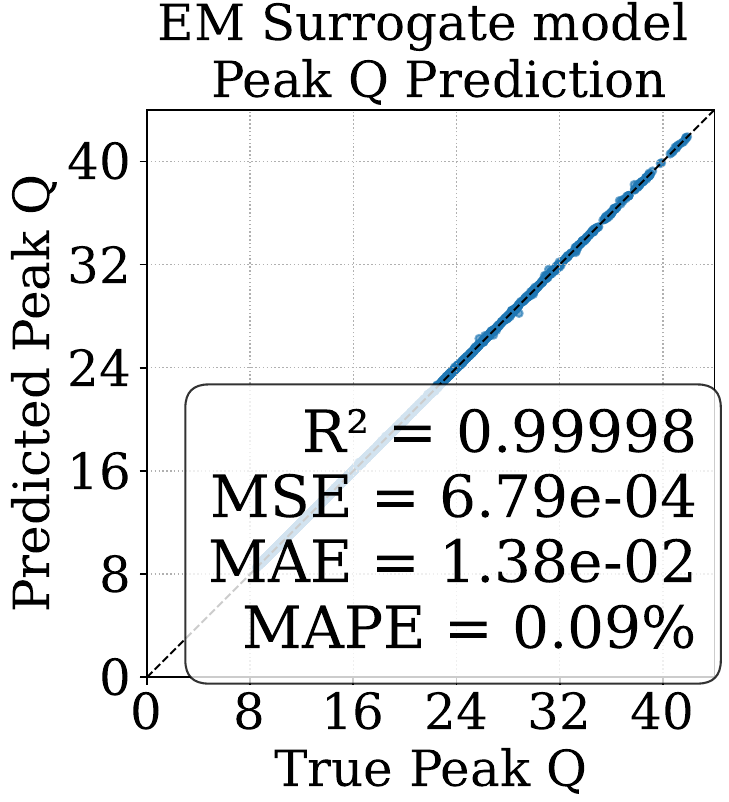} &
        \includegraphics[width=0.26\linewidth]{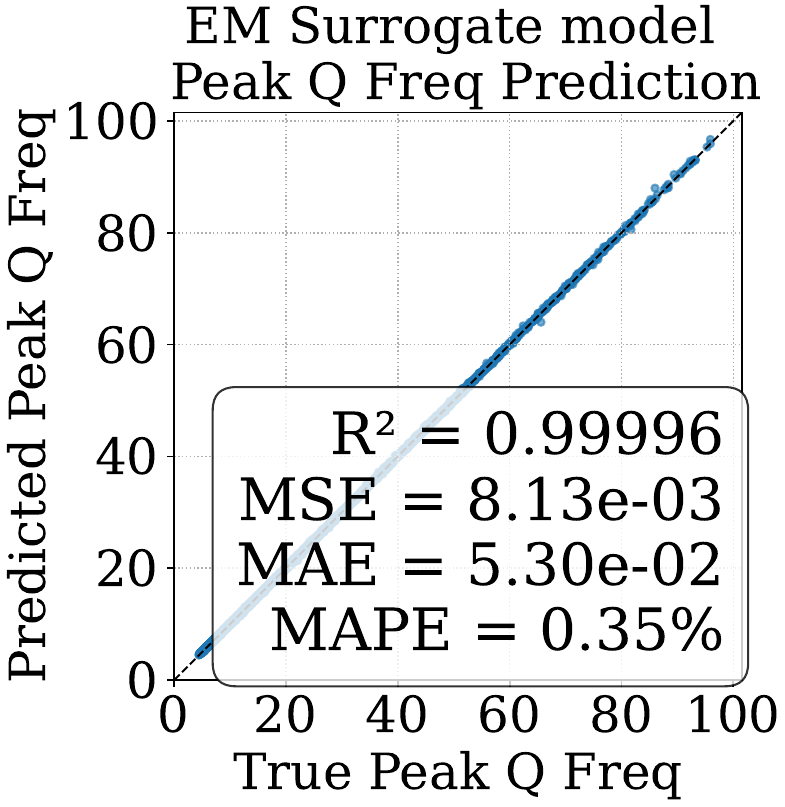} &
        \includegraphics[width=0.25\linewidth]{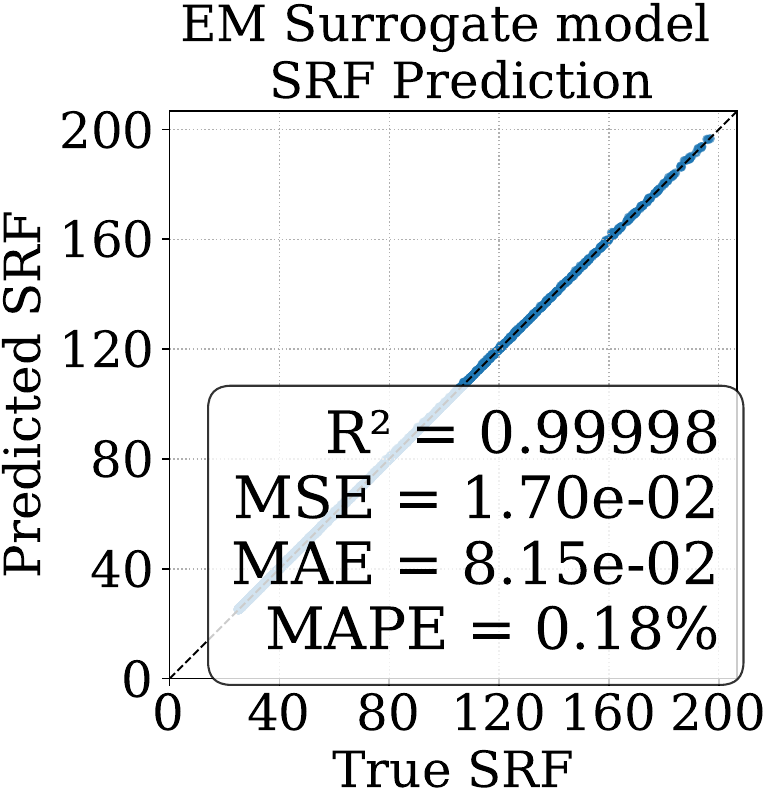} \\
        (a) & (b) & (c) & (d)
    \end{tabular}
    \caption{Scatter plots of (a) $L$, (b) $Q_{peak}$, (c) $f_{peakQ}$, and (d) $f_{SRF}$.}
    \Description{Scatter plots of (a) $L$, (b) $Q_{peak}$, (c) $f_{peakQ}$, and (d) $f_{SRF}$.}
    \label{fig:emx_surrogate}
    \vspace{-2mm}
\end{figure}

For \textbf{spiral inductors}, a forward random-forest~\cite{random_forest} EM model is combined with an inverse k-NN-based synthesizer~\cite{knn}.
The \textit{forward EM model} learns the mapping from geometry parameters $\mathbf{X}_L = [t, w, s, r]$ to the EM response, where $t$ is the number of turns, $w$ the metal width, $s$ the spacing between turns, and $r$ the inner radius. A dataset of 80k inductors,
laid out using a Python layout generator, was simulated in Cadence EMX over 24~hours. 
An ensemble random-forest model with 200 estimators, trained with mean-squared error loss and bootstrap aggregation, predicts $\mathbf{Y_{EM}}=(L,\,Q_{peak},\,f_{peakQ},\,f_{SRF})$ with high accuracy (R$^2$$>$0.999, Fig.~\ref{fig:emx_surrogate}). 

\begin{figure}[b]
    \centering
    \setlength{\tabcolsep}{0pt} 
    \renewcommand{\arraystretch}{0}
    \begin{tabular}{cccc}
        \includegraphics[width=0.17\linewidth]{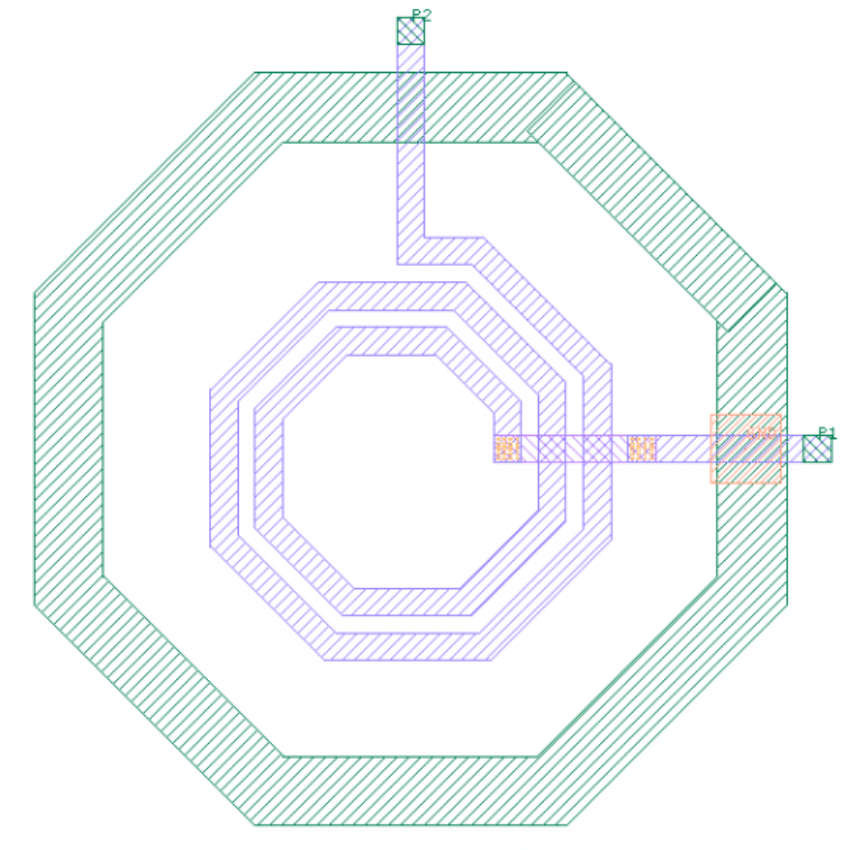} \hspace{4mm} &
        \includegraphics[width=0.19\linewidth]{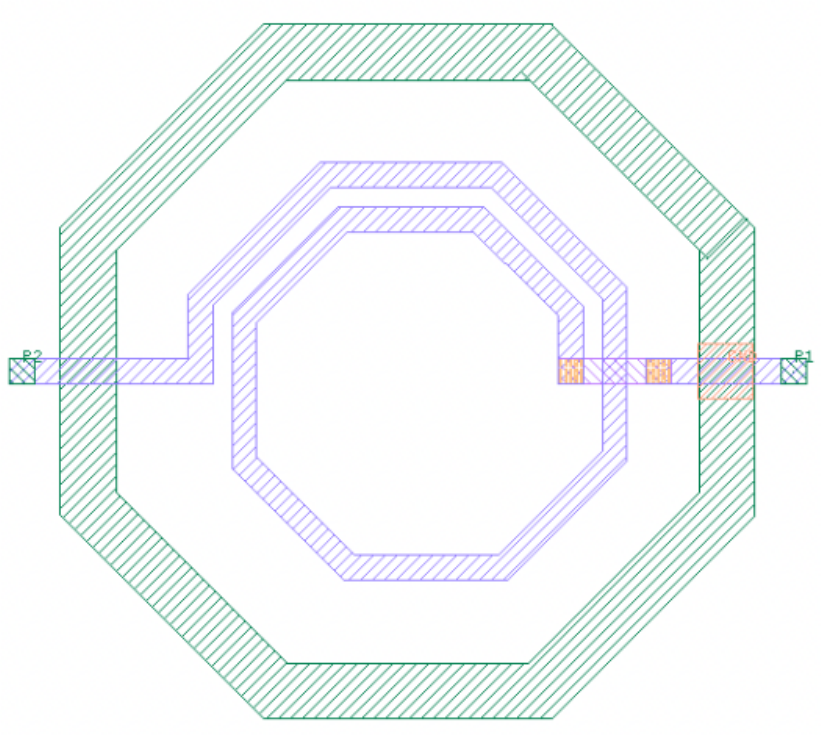}  \hspace{4mm} &
        \includegraphics[width=0.17\linewidth]{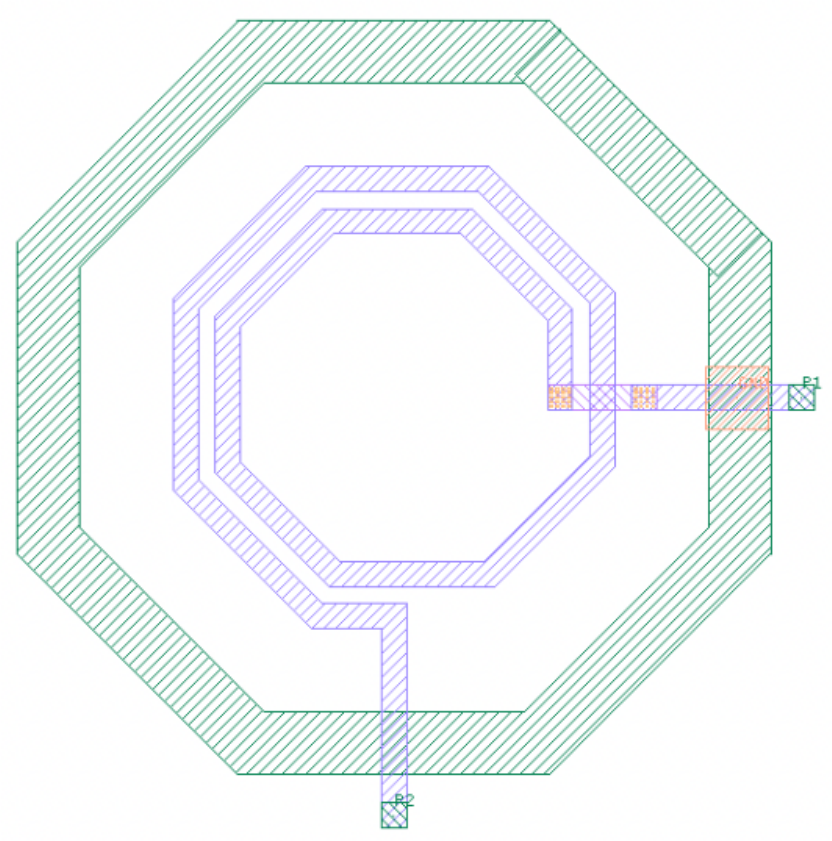} \hspace{4mm} &
        \includegraphics[width=0.18\linewidth]{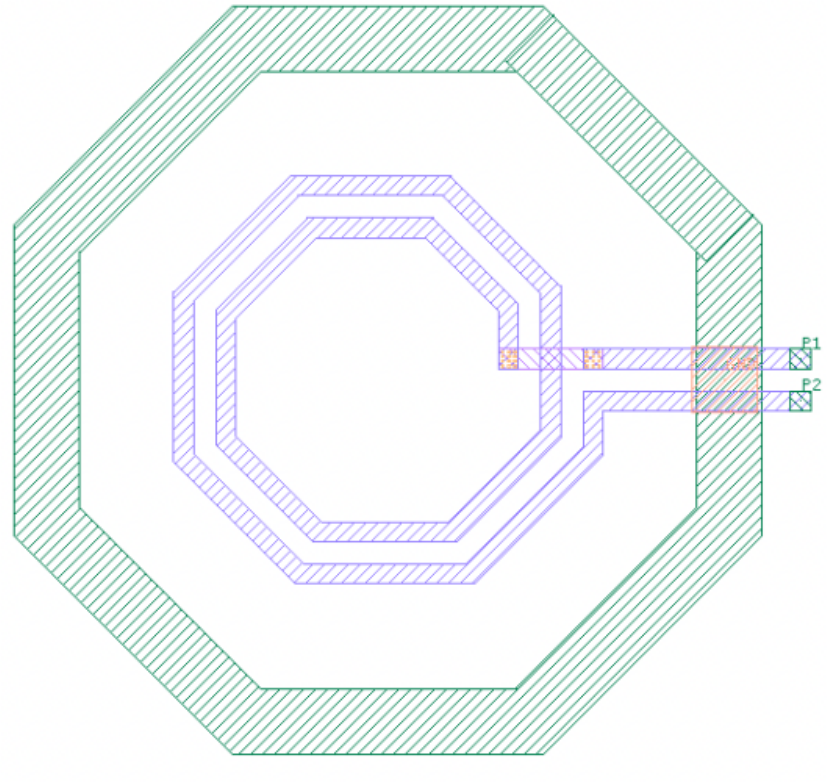} \\
        \hspace*{-5mm} {\small (a) $\tfrac{1}{4}$-turn ($\tfrac{1}{4}$T)} & 
        \hspace*{-5mm} {\small (b) $\tfrac{1}{2}$-turn ($\tfrac{1}{2}$T)} & 
        \hspace*{-5mm} {\small (c) $\tfrac{3}{4}$-turn ($\tfrac{3}{4}$T)} & 
        {\small (d) 1-turn (1T)}
    \end{tabular}
    \caption{Fractional-turn variants with similar inductance.}
    \Description{Fractional-turn variants with similar inductance.}
    \label{fig:inductor_variants}
\end{figure}
 
The \textit{inverse k-NN-based synthesizer} identifies parameters $\mathbf{X}_L=[t,w,r,s]$ that meet desired EM behavior. Given an $L_{target}$ from the matching network, the optimal choice among candidates $\mathbf{X}_L$ with similar inductance (within $\epsilon = 5\%$ value), maximizing:

\vspace{-5mm}
\begin{equation}
\textstyle \Big( \sum_{i\in\mathcal{M}} Y_{EM,i}(\mathbf{X}_L) \Big)
-
\Big( \sum_{i\in\mathcal{F}} (Y_{EM,i}(\mathbf{X}_L)-Y_{EM,i,\text{target}})^2 \Big),
\label{eq:Stage1_objective}
\end{equation}
\vspace{-3mm}

\noindent
where $Y_{EM,i}$ represents a performance parameter evaluated by the forward EM model. Some parameters $\mathcal{F}$ must be fixed ($L_{target}$) within a tolerance, while others are maximized ($Q_{peak}, f_{peakQ}, f_{SRF}$).

\begin{figure}[t]
    \centering
    \includegraphics[width=0.8\linewidth]{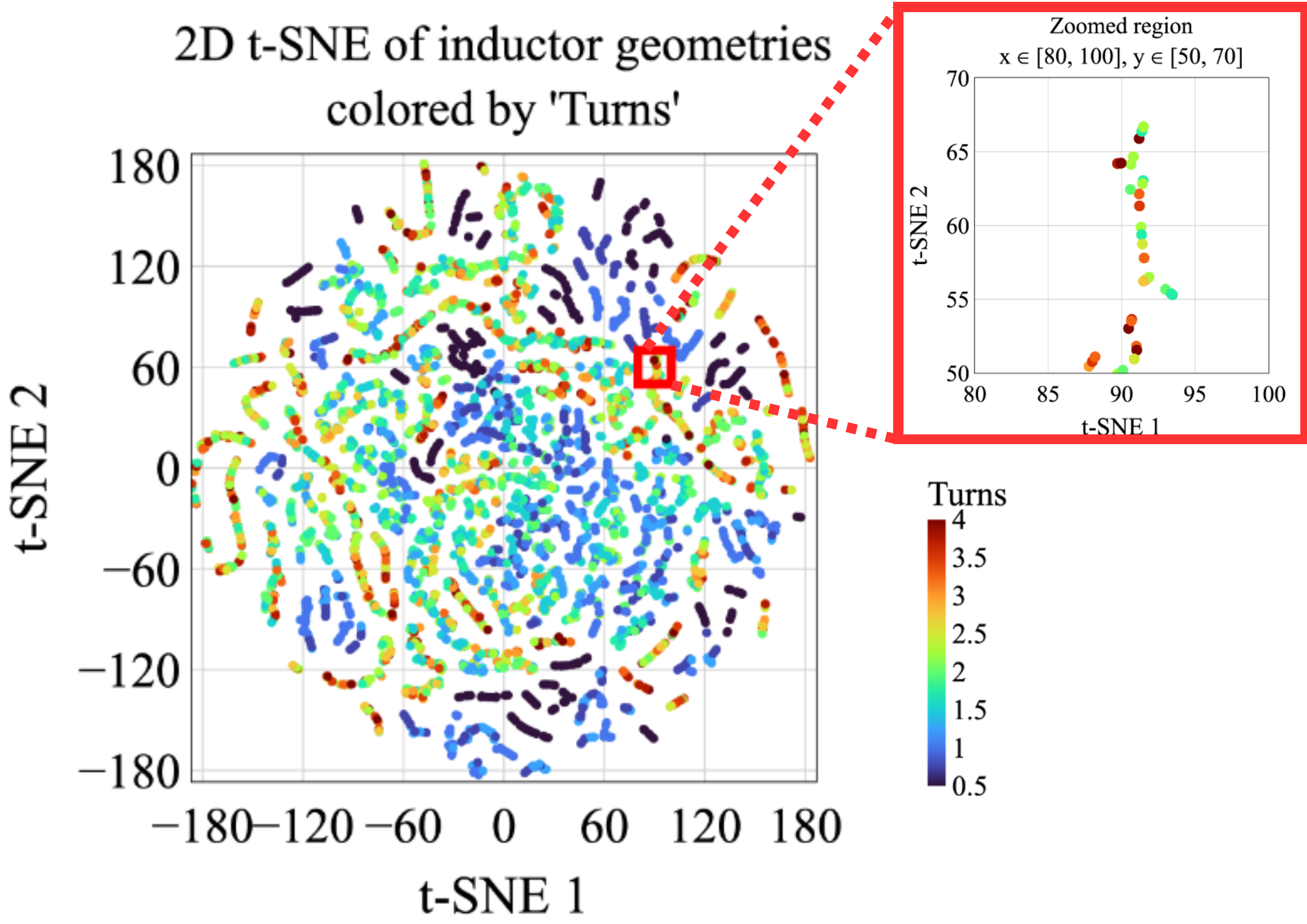}
    \caption{2D t-SNE visualization of inductor geometry clusters and a zoomed-in cluster showing EM-equivalent variants.}
    \Description{2D t-SNE visualization of inductor geometry clusters and a zoomed-in cluster showing EM-equivalent variants.}
    \label{fig:tsne_cluster}
\end{figure}

At the layout stage, it is important to obtain a diversity of solutions with various locations for the terminals of the inductor. Fig.~\ref{fig:inductor_variants} illustrates layouts with four combinations of terminal locations, referred to as $\tfrac{1}{4}$T, $\tfrac{1}{2}$T, $\tfrac{3}{4}$T, and 1T variants. Note that since inductor dimensions are large, this flexibility is vital: for example, a 1T solution incurs prohibitive parasitics to connect to a block at the opposite end of the inductor; however, these parasitics are already accounted for in a $\tfrac{1}{2}$T layout. Each node in the dataset is labeled according to its terminal locations, and we can select four solutions from the k-NN, maximizing~\eqref{eq:Stage1_objective} for each fractional turn $f \in \left \{ \tfrac{1}{4}\text{T}, \tfrac{1}{2}\text{T}, \tfrac{3}{4}\text{T}, 1\text{T} \right \}$. Fig.~\ref{fig:tsne_cluster} shows distinct clusters of geometries with similar inductances but varying $Q_{peak}$ and $SRF$, enabling selection of the most efficient variant for each $L_{target}$.

For \textbf{CPWs} and \textbf{T-lines}, a single k-NN regressor is used in the neighborhood of the specification to infer design parameters.  The regression provides physical parameters, $\mathbf{X}_{CPW} = [ l, w, s ]$ and $\mathbf{X}_{TL} = [ l, w ]$, where $l$, $w$, and $s$ are the length, width, and spacing.  
The map from geometry to parameters is lower-dimensional and smoother than in inductors, and an EM surrogate is not needed.

For passive flexibility, COmPOSER selects between T-lines and spiral inductors based on $L_{target}$. Values below $L_{\min}$ use T-lines due to their smaller area and improved performance, whereas larger values requiring one or more full turns are implemented as spiral inductors. In the 65~nm technology that we use, $L_{\min}=100$\,pH.

\noindent
\textbf{Hybrid analytical sizing.}
COmPOSER analytically optimizes LNA and PA candidates by coupling closed-form circuit equations with ML-based EM refinement, bridging schematic-level estimation with layout-consistent accuracy.
All equations and symbols are defined in Table~\ref{tab:analytical_eqs}:
\eqref{eq:nf_min}--\eqref{eq:lna_eqs} capture LNA performance behavior; \eqref{eq:pa_stages}--\eqref{eq:Z_l_gate} describe PA load-line and power-transfer relations; and \eqref{eq:L_omega} tunes matching networks for both~\cite{razavi, sorin_book, Lee_2003}. 
Layout-dependent parasitics for active devices, captured through PDK-provided models for $C_{gs}$, $C_d$, $R_g$, and $r_o$, are parameterized by the number of fingers, $N_f$, enabling accurate prediction of capacitance scaling and gate-resistance reduction without full layout. Passive elements are initially modeled as ideal lumped $L$–$C$ components and subsequently corrected by the ML-based EM optimizer. \textit{Section~\ref{sec:exp_results}
demonstrates that the hybrid estimator--arising from analytical device models augmented by EM-calibrated passive models--is sufficiently accurate for meeting mm-wave/RF LNAs and PAs specifications.}

\begin{table}[t]
\scriptsize
\centering
\caption{Design-specific equations \ignore{in COmPOSER} for LNA and PA.}
\label{tab:analytical_eqs}

\renewcommand{\arraystretch}{1.15}
\setlength{\tabcolsep}{1pt}

\begin{tabular}{|w{c}{0.3cm}|p{0.95\columnwidth}|}
\hline
\multicolumn{2}{|c|}{\textbf{LNA equations (gain, noise, and matching analysis)}} \\
\hline  

\eqLabel{eq:nf_min} &
$\displaystyle
\begin{array}{l}
N = (f_T / (f_0 R_{in})) \sqrt{R_g / (\gamma g_m (4 / ((f_T/f_0)^2 + 1) + 1))}
\end{array}
$
\\
\hline

\eqLabel{eq:L_sdg} &
$\displaystyle
\begin{array}{l}
L_s = f(R_{in},R_g,r_o,(C_{gd}/C_{gs}),N); \quad
L_d = 1 / ((N C_d + C_{load}) \omega_0^2); \\
L_g = 1 / (N C_{gs} \omega_0^2) - L_s;
\end{array}
$
\\
\hline

\eqLabel{eq:Z_in_1} &
$\displaystyle
\begin{array}{l}
Z_{in,MOS} = R_{in} + j(L_s \omega_0 - 1 / (N C_{gs} \omega_0))
\end{array}
$
\\
\hline

\eqLabel{eq:Z_in_2} &
$\displaystyle
\begin{array}{l}
Z_{in,LNA} =
Z_{cpw}
\frac{
Z_g((Z_{in,MOS}+Z_g\tan\theta_g)/(Z_g+Z_{in,MOS}\tan\theta_g)) + Z_{cpw}\tan\theta_{cpw}
}{
Z_g((Z_{in,MOS}+Z_g\tan\theta_g)/(Z_g+Z_{in,MOS}\tan\theta_g))\tan\theta_{cpw} + Z_{cpw}
}
\end{array}
$
\\
\hline

\eqLabel{eq:lna_eqs} &
$\displaystyle
\begin{array}{l}
S_{11} = 20\log_{10}(|(Z_{in,LNA}-Z_0)/(Z_{in,LNA}+Z_0)|); \quad
G = (R_d / (4 R_{in})) (f_T/f_0)^2; \\
NF = 1 + \gamma g_m N R_{in}(f_0/f_T)^2(4/((f_T/f_0)^2+1)+1) 
+ 4R_{in}(f_0/f_T)^2/R_d~+ \\ 
\hspace*{12mm} (R_g/N + R_{loss})/R_{in}; \quad
BW = f_0 / Q_{out}; P = N P_{unit};
\end{array}
$
\\
\hline
\hline

\multicolumn{2}{|c|}{\textbf{PA equations (load-line and power relations)}} \\
\hline

\eqLabel{eq:pa_stages} &
$\displaystyle
\begin{array}{l}
N_{stages} =
\lceil G / (10\log_{10}((R_d/(4R_g))(f_T/f_0)^2)) \rceil
\end{array}
$
\\
\hline

\eqLabel{eq:OP} &
$\displaystyle
\begin{array}{l}
1 / P_{sat,out}
= \sum_{i=1}^{N}
1 / (P_{sat,i} \prod_{k=i+1}^{N} G_i); \quad
P_{sat,i} = W_i J_{opt} (V_{DD}-V_{DSAT}) / 4;
\end{array}
$
\\
\hline

\eqLabel{eq:width_of_stage_i} &
$\displaystyle
\begin{array}{l}
W_i = W_{i+1} (W_{in}/W_{out})^{1/(N_{stages}-1)}
\end{array}
$
\\
\hline

\eqLabel{eq:Z_l_gate} &
$\displaystyle
\begin{array}{l}
Z_{i,drain} =
1 / (W_iJ_{opt}/(2(V_{DD}-V_{DSAT})) + j\omega_0 W_i C_d); \\
Z_{i+1,gate} =
R_g/W_{i+1} + 1 / (j\omega_0 W_{i+1} C_{gs});
\end{array}
$
\\
\hline
\hline

\multicolumn{2}{|c|}{\textbf{Shared matching equation (LNA and PA)}} \\
\hline

\eqLabel{eq:L_omega} &
$\displaystyle
\begin{array}{l}
Z = 2\pi f_0 L / \tan(2\pi f_0 / (4 f_{SRF}))
\end{array}
$
\\
\hline
\hline

\multicolumn{2}{|c|}{\textbf{List of symbols used} (other symbols defined in Sections~\ref{sec:Intro},~\ref{sec:man_design}, and ~\ref{sec:stage_1})} \\
\hline
\multicolumn{2}{|p{0.98\columnwidth}|}{
\scriptsize
\renewcommand{\arraystretch}{1.05}
\begin{tabular}{@{}p{1.3cm}p{3.2cm}|p{0.87cm}p{3.2cm}@{}}

$N$         & MOS width multiplier
& $f_0, \omega_0$  & Operating frequency; $\omega_0 = 2\pi f_0$ \\

$R_{in}, R_{g}, R_{d}$         & MOS input, gate, and drain resistance   
& $g_m$          & MOS transconductance \\

$\gamma$       & Thermal-noise coefficient 
& $r_o$          & Output resistance \\

$C_{gs}, C_{gd}, C_d$  & MOS intrinsic and parasitic caps    
& $C_{load}$     & Load capacitance \\

$Z_{in,MOS}$     & MOS gate input impedance         
& $Z_{in,LNA}$   & LNA input impedance \\

$Z_g, \theta_g$  & Impedance and phase of gate T-line 
& $S_{11}$         & Input reflection coefficient   \\

$Z_0$            & Reference impedance (50~$\Omega$) 
&$R_{loss}$     & Passive-path loss resistance \\

$Q_{out}$        & Output network quality factor       
& $P_{unit}$       & Power of a unit device  \\

$N_{stages}$   & Number of PA stages 
& $P_{sat,out}$  & Output-stage saturation power \\

$G_i$          & Gain of stage $i$ 
&$P_{sat,i}$      & Saturation power of stage $i$  \\

$V_{DSAT}, V_{DD}$       & Saturation voltage, supply voltage 
& $W_i$            & Width of stage $i$  \\                       
                       
$Z_{i,drain}$  & Drain impedance of stage $i$     
& $Z_{i,gate}$   & Gate impedance of stage $i$ \\

\end{tabular}
}
\\
\hline

\end{tabular}
\end{table}

\begin{algorithm}[t]
\footnotesize
\caption{Analytical estimation of LNA and PA candidates}
\label{algo:amp}
\begin{algorithmic}[1]

\State \textbf{Input:} Specifications on $\mathcal{S} = \{f_0,\, G,\, NF,\, BW,\, C_{load},\, P_{sat},\, S_{11, max}\}$ 
\label{step:InputSpecs}

\State \textbf{Technology parameters:} $g_m,\, C_{gs},\, C_d,\, R_g,\, f_T,\, V_{DSAT},\, \gamma, W_{in}$
\label{step:TechParams}

\State \textbf{Output:} Active and passive sizes
\label{step:Output}
\vspace{3pt}

\If{Topology = LNA}
    \State $\mathcal{M}$ from Equation~\eqref{eq:nf_min} and $NF$--$BW$--$G$ tradeoffs~\cite{sorin_2007, razavi}
    
    \label{step:LNA_Ni}

    \ForAll{$N \in \mathcal{M},\; R_{in} \in [40,60]~\Omega$}
    \label{step:lna_start}
        \State Evaluate $(L_s, L_g, L_d)$
        \Comment{Equation~(\ref{eq:L_sdg})}
        \label{step:lna_match}

        \For{$k \in \{s, g, d\}$}
        \Comment{Section~\ref{sec:stage_1}}
        \label{step:lna_emloop}
            \State $\mathbf{Y_{EM}}
            \!\leftarrow\!$ \textit{ML–based EM optimizer}$(L_k)$
            \label{step:lna_emopt}

            \State Update $(C_{in},\, C_d,\, C_{out}) \!\leftarrow\! f(\mathbf{Y_{EM}})$
            \Comment{Equation~(\ref{eq:L_omega})}
            \label{step:lna_cap_update}
            
        \EndFor\vspace{-2pt}
        \ForAll{$Z_{cpw}\!\in\![Z_{min},Z_{max}]~\Omega,\, \theta_{cpw}\!\in\![\theta_{min},\theta_{max}]~\text{rad}$}
        \label{step:lna_cpw_sweep}

        \State Evaluate $Z_{in,mos}, Z_{in,LNA}, S_{11}$ 
        \Comment{Equations~(\ref{eq:Z_in_1}),(\ref{eq:Z_in_2}),(\ref{eq:lna_eqs})}
        \label{step:lna_cpw_zin}
        
        \State \textbf{if} $S_{11}\!\le\!S_{11,max}$ \textbf{then}
        \label{step:lna_cpw_check}
        \State\hspace{1.5em}$\text{Dimension of CPW}\!\leftarrow\!\textit{CPW\_k-NN}(Z_{cpw},\theta_{cpw})$
        \Comment{Section~\ref{sec:stage_2}}
        \label{step:lna_cpw_map}
        
        \EndFor\vspace{-2pt}
        \State Evaluate $\{NF,\, BW,\, G\}$
        \Comment{Equation~(\ref{eq:lna_eqs})}
        \label{step:lna_eval}
    \EndFor\vspace{-2pt}
    \label{step:lna_end}

    \State Select design with minimum deviation from $\mathcal{S}$
    \label{step:lna_select}
    
\ElsIf{Topology = PA}
    \State Calculate $N_{stages}$
    \Comment{Equation~(\ref{eq:pa_stages})}
    \label{step:PA_stage_scale}
    \State Derive $W_{out}$ from $P_{sat}$ and $N_{stages}$
    \label{step:PA_scale}
    \Comment{Equation~(\ref{eq:OP})}
    
    \ForAll{$i = (N_{stages}-1) \to 1$}
    \label{step:pa_loop}
        \State Calculate $W_i$ 
        \Comment{Equation~\eqref{eq:width_of_stage_i}}
        \label{step:size_of_i}

        \State Matching network $\leftarrow f(W_{i+1}, W_i)$
        \Comment{Equation~(\ref{eq:Z_l_gate})}
        \label{step:pa_match}

        \For{$k \in \{d, g\}$}  
        \Comment{Section~\ref{sec:stage_1}}
        \label{step:pa_emloop}
            \State \hspace{-0.2em}$\mathbf{Y_{EM}}\!\leftarrow\!\textit{ML–based EM optimizer}(L_k)$
            \label{step:pa_emopt}

            \State Update $(C_{series},\, C_{d}) \!\leftarrow\! f(\mathbf{Y_{EM}})$
            \Comment{Equation~(\ref{eq:L_omega})}
            \label{step:pa_cap_update}

        \EndFor
    \EndFor
    \label{step:pa_loopend}
\EndIf\vspace{-2pt}
\label{step:end_if}
\State \textbf{Return:} Active and passive sizes
\label{step:Return}

\end{algorithmic}
\end{algorithm}

Algorithm~\ref{algo:amp} defines the analytical sizing flow. 
For the LNA, the feasible cascode transistor sizes $N \in \mathcal{M}$ are obtained (line~\ref{step:LNA_Ni}) from the minimum-$NF$ relation in Eq.~\eqref{eq:nf_min} together with the $NF$--$BW$--$G$ trade-offs in~\cite{sorin_2007, razavi}. Each pairwise tradeoff results in a quadratic, whose two roots provide a set of candidate sizes, which are combined to populate a small set $\mathcal{M}$ that sharply prunes the search space.
The allowable input-impedance range is restricted to $R_{in}\!\in\![40,60]~\Omega$ to cover realizable impedances around the ideal $50~\Omega$ while accounting for post-layout parasitics.  Line~\ref{step:lna_start} begins a loop that sweeps the $(N, R_{in})$ pairs. Within this loop, lines~\ref{step:lna_match}--\ref{step:lna_cap_update} compute $L_s$, $L_g$, and $L_d$ using Eq.~\eqref{eq:L_sdg}; evaluate the inductors EM behavior using the ML-based EM optimizer; and update $(C_{in}, C_d, C_{out})$ through Eq.~\eqref{eq:L_omega}$,$ ensuring that the sizing accounts for device parasitics.
Line~\ref{step:lna_cpw_sweep} sweeps the CPW parameters\ignore{$(Z_{cpw},\theta_{cpw})$} for input matching over $Z_{cpw}\in[Z_{min},Z_{max}]$ and $\theta_{cpw}\in[\theta_{\min},\theta_{\max}]$ (where $Z_{min}=18~\Omega$, $Z_{max}=50~\Omega$, $\theta_{\min}=0$, $\theta_{\max}=\pi/2$) for the 65~nm CMOS stack.
Candidates meeting the $|S_{11}|$ target are mapped to geometries using the k-NN-based CPW model. 
\ignore{Finally, }$NF$, $BW$, and $G$ are then evaluated (Eq.~\eqref{eq:lna_eqs}), and the best configuration is selected in line~\ref{step:lna_select}.

For the PA, line~\ref{step:PA_stage_scale} uses $G$ to determine $N_{stages}$, the number of cascaded stages using Eq.~\eqref{eq:pa_stages}, and line~\ref{step:PA_scale} estimates the output stage width, $W_{out}$, from $P_{sat}$ using Eq.\eqref{eq:OP}. The input stage size, $W_{in}$, corresponds to the device size delivering roughly 0~dBm; in 65~nm, $W_{in}=8\mu$m. Line~\ref{step:pa_loop} then iterates backward from output to input, computing each preceding width $W_i$ (line~\ref{step:size_of_i}) using a geometric scale.
The inter-stage matching network is generated from the adjacent stage sizes $(W_{i+1}, W_i)$ using Eq.~\eqref{eq:Z_l_gate} (line~\ref{step:pa_match}), and the inductors $L_d$ and $L_g$ are evaluated using the ML-based EM optimizer in lines~\ref{step:pa_emloop}–\ref{step:pa_emopt}.
The resulting EM responses $\mathbf{Y_{EM}}$ are used to update $C_{series}$ and $C_d$ via Eq.~\eqref{eq:L_omega}. Algorithm~\ref{algo:amp} thus returns EM-consistent device sizes, enabling 
EM-calibrated sizing.
\vspace{-2mm}

\vspace{-6mm}
\section{Stage 2: Primitive Block Layout Generator}
\label{sec:stage_2}

\vspace{-3pt}
\noindent
Stage~2 generates layout options for the building blocks, or primitives, of the optimized circuit from Stage~1. Primitive layout generators translate parameters from Stage~1
to DRC-clean layouts.

\begin{figure}[h]
    \centering
    \setlength{\tabcolsep}{0pt} 
    \renewcommand{\arraystretch}{0} 

    \subfloat[ ]{%
        \includegraphics[width=0.17\linewidth]{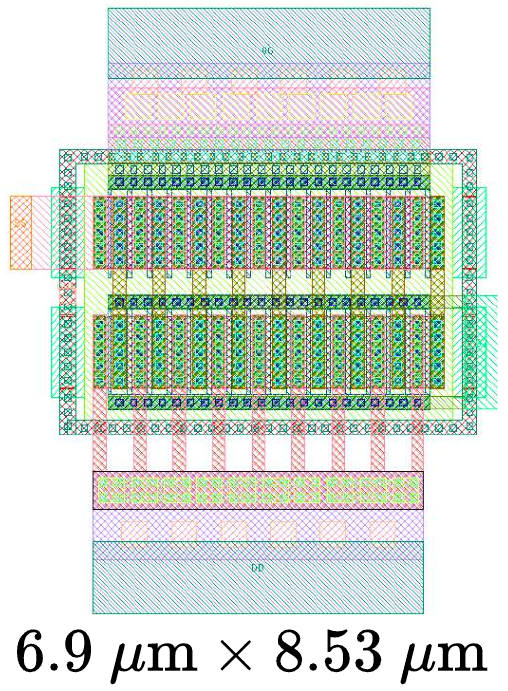}%
        \label{fig:primitives_a}
    }\hspace{1mm}
    \subfloat[]{%
        \includegraphics[height=0.18\linewidth]{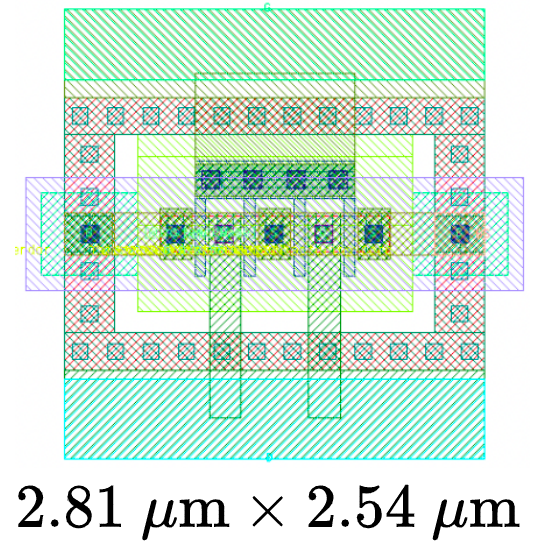}%
        \label{fig:primitives_b}
    }\hspace{1mm}
    \subfloat[]{%
            \includegraphics[height=0.22\linewidth]{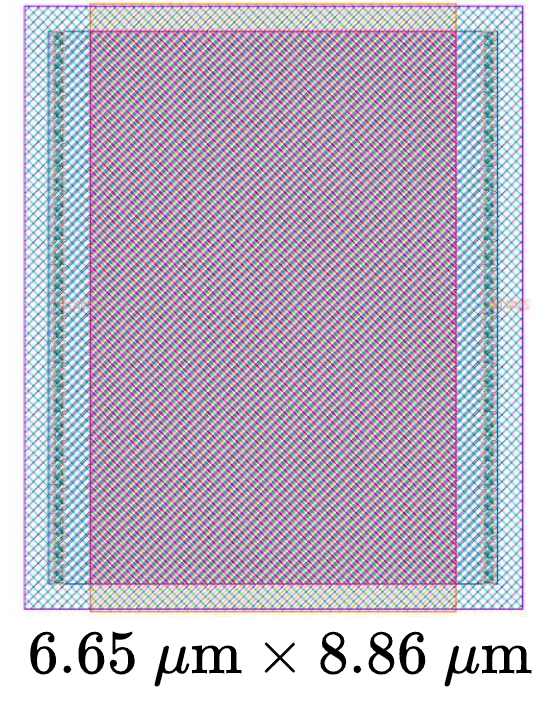}%
        \label{fig:primitives_c}
    }\hspace{1mm}
    \subfloat[]{%
        \includegraphics[width=0.14\linewidth]{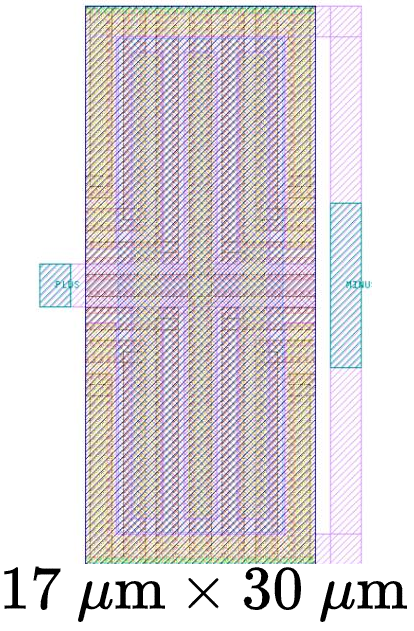}%
        \label{fig:primitives_d}
    }\hspace{1mm}
    \subfloat[]{%
        \includegraphics[height=0.23\linewidth]{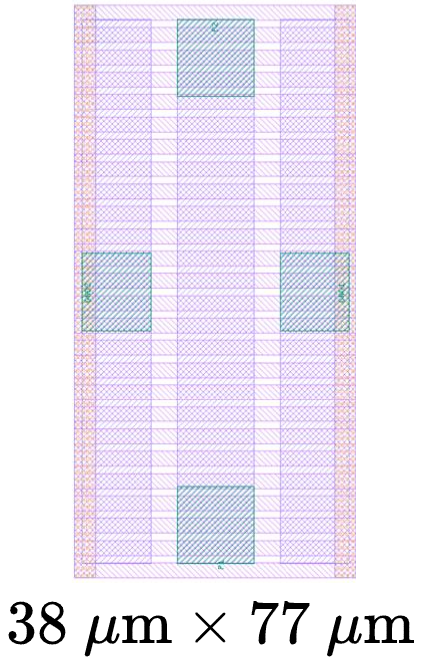}%
        \label{fig:primitives_e}
    }
     \vspace{-6pt}

    \caption{Layout primitives \textit{(not to scale)} for 
    (a) Cascode MOS, (b) Biasing MOS, (c) Resistor, (d) Capacitor, and
    (e) CPW.}
    \Description{Layout primitives \textit{(not to scale)} for 
    (a) Cascode MOS, (b) Biasing MOS, (c) Resistor, (d) Capacitor, and
    (e) CPW.}
    \label{fig:primitives}
    \vspace{-2mm}
\end{figure}

\noindent
\textbf{Active devices.}
Active device generators create cascode and biasing MOS primitives using the transistor dimensions---number of fingers ($N_f$), finger width ($W_f$), and channel length ($L$)---obtained in Stage 1. In cascode configurations, the transistors utilize a staircase layout structure along the RF path, adopted from~\cite{staircase_MOS}; this topology minimizes intermediate parasitic capacitance and enhances high-frequency gain. Representative layouts are shown in Fig.~\ref{fig:primitives}(a)–(b).

\noindent
\textbf{Resistors and capacitors.}  
Resistors (Fig.~\ref{fig:primitives}(c)) are modeled as $R = R_s \tfrac{l}{w}$, where $R_s$ is the sheet resistance, with multiple $(l/w)$ variants generated per target to trade off area and routing flexibility. MIM capacitors (Fig.~\ref{fig:primitives}(d)) follow a PDK-calibrated model $C_{eff}(l,w) = \alpha_l l + \alpha_w w + \alpha_{lw}lw + \alpha_0$, capturing both parallel-plate and fringing effects. Aspect-ratio variants of both devices are generated.

\noindent
\textbf{Inductors and CPWs.} For spiral inductors parameterized by $\mathbf{X}=[t,\, r,\, w,\, s]$, the primitive generator builds DRC-clean fractional-turn layout variants. These EM-equivalent alternatives allow the placer to balance area and routing length by selecting geometries with comparable inductance but differing footprints and pin locations.  T-line inductors with parameters $[l, w]$ and CPWs (Fig.~\ref{fig:primitives}(e)) with parameters $[l, w, s]$ are also translated to DRC-clean layouts.
\vspace{-2mm}
\section{Stages 3, 4: Layout Synthesis and Verification}
\label{sec:stage_3}

\noindent
\textbf{Stage~3} performs physical layout synthesis by hierarchically placing primitives, routing interconnects, and constructing the PDN with decap integration, incorporating any user-provided constraints such as labeled critical nets, PDN density, and halo spacing.

\noindent
\textbf{MILP-based placement.} Placement is formulated as a 0-1 MILP:

\vspace{-3mm}
{\small
\begin{equation*}
\begin{alignedat}{2}
\textstyle \min_{\mathbf{x},\mathbf{y},\mathbf{r},\mathbf{s}}~~
&[(x_n^{\max}-x_n^{\min})+(y_n^{\max}-y_n^{\min})]
+ \textstyle \sum_{n\in\text{nets}}\omega_n\,\mathrm{HPWL}(n) \\[-2pt]
\text{s.t.}\quad
& (1) \textstyle \sum_{r} r_{i,j}=1,\ \forall i\in\mathcal{M}, \qquad
(2) \sum_{j} s_{i,j}=1,\ \forall i\in\mathcal{M}, \\[-2pt]
& (3)~\text{Big-}M\ \text{non-overlap on }(w_i{+}2d_{\mathrm{halo}},h_i{+}2d_{\mathrm{halo}}),
&&\forall i\neq j,\\[-2pt]
&(4)~u_{i,N}+u_{i,E}+u_{i,S}+u_{i,W}=1, &&\forall i\in\mathcal{P},\\[-2pt]
&(5)~x_i=k_i p_{\mathrm{IO}}\ \text{or}\ y_i=k_i p_{\mathrm{IO}}, &&\forall i\in\mathcal{P}.
\end{alignedat}
\end{equation*}
}

\vspace{-2mm}

The first term of the objective minimizes area and the second the weighted half-perimeter wirelength (HPWL)~\cite{hpwl}; 
per-net weights $\omega_n$ scale HPWL, with larger weights on user-specified RF-critical nets (red nets in Fig.~\ref{fig:lna_pa}), and noncritical net weights set to 1. The placer optimizes module coordinates, and selects among block variants and rotations, under nonoverlap, halo, and I/O pad constraints. 

For a set of modules $\mathcal{M}$ and I/O pads $\mathcal{P}\subset\mathcal{M}$, each module $i$ is represented by continuous coordinates $(x_i,y_i)$ with up to four one-hot binary variables, $r_{i,j}$, corresponding to rotations \{0, 90$^\circ$, 180$^\circ$, 270$^\circ$\}; variables for forbidden rotations (e.g., \{90$^\circ$, 270$^\circ$\} for transistors) are omitted. Modules with $V_i$ variants use one-hot binary variables $s_{i,1},\dots,s_{i,V_i}$; 
Constraints~(1) and~(2) ensure that exactly one variant and one rotation is chosen. Rotation and variant selections are translated to effective $x$ and $y$ block dimensions through auxiliary variables. Overlap  is avoided in constraint~(3) through four-way Big-$M$ disjunction~\cite{wolsey_integer_programming_2020}. Traditional big-M constraints~\cite{Kim03} are extended to footprints expanded by the user-defined halo $d_{\mathrm{halo}}$ between
modules, thus reserving routing corridors. These corridors are especially important near large RF passives (e.g., inductors), which are routing blockages, prohibiting above- or below-the-block routing. For pads $i\in\mathcal{P}$, one-hot indicators $u_{i,N},u_{i,E},u_{i,S},u_{i,W}$ restrict placement to the allowed chip edges (North, East, South, and West, respectively) in Constraint~(4); integer variables $k_i$ snap pad coordinates to the DRC-specified I/O pitch $p_{\mathrm{IO}}$ ($\sim$$\mu$m) in Constraint~(5). Internal modules are placed on a much finer DRC grid and can be continuously placed in practice.

Gurobi~\cite{gurobi} solves the MILP with a $10^{-3}$ MIP gap, selecting
variants, orientations, and pad-side locations to produce compact,
DRC-compliant floorplans. Typical designs (28--30 modules) yield $\sim$2.3K
variables and $\sim$3.3K constraints, and are solved within 20~s.

\noindent
\textbf{A$^*$ routing.}
The router performs EM-consistent interconnect synthesis, constructing a Hanan-based Manhattan grid from pin locations, blockage boundaries, and existing metals.
Each module blockage is inflated by $(w_{\text{route}}/2 + s_{\min})$ on each side to obey DRCs between routed wires and internal within-module wires.
Note that this expansion is 
separate from the user-defined placement halo that creates routing channels around blocks during placement.

The router supports bidirectional routing
while respecting preferred layer orientations in the PDK.  A minimum spanning tree is built for each net and used to decompose the net into pairwise connections to build Steiner trees.  Paths are computed using an A$^*$ search~\cite{hart_astar_1968}, where edge weights combine interconnect/via resistance and net priority. Critical nets are routed on upper metal layers with increased width to reduce loss.
Net importance guides width and layer assignment and weights the routing cost in the A$^*$ search.  
For our LNA and PA testcases, routing completes in under five seconds and scales approximately linearly with grid size.

\noindent
\textbf{PDN synthesis and decap placement.}
Unlike prior approaches, COmPOSER integrates PDN synthesis into the flow, using user-specified metal allocations, typically M1--M6 for $GND$ and M7--M9 for $V_{DD}$, together with allowable strap widths and thicknesses. A DRC-clean mesh is generated while excluding regions above or below shielded passives (e.g., inductors and T-lines).
Precharacterized 500~fF decaps are inserted across all legal whitespace, yielding a supply network with inherently low impedance. Although PDN density can affect stability, we provision a sufficiently dense grid and aggressive decap coverage from the outset 
to avoid adaptive refinement. 
The full procedure completes in under 20~s. As shown in Section~\ref{sec:exp_results}, our method consistently drives all designs from $\kappa<1$ to $\kappa>1$, where $\kappa$ is the Rollett stability factor~\cite{rollett_1962} certifying unconditional stability, 
showing the effectiveness of COmPOSER.

\noindent
\textbf{Stage~4} of COmPOSER performs a full post-layout verification via EM–circuit co-simulation. The finalized layout undergoes full-wave EM analysis using Cadence EMX to extract frequency-dependent parasitics, capturing coupling, current crowding, and substrate loss. These parasitics are back-annotated into the netlist for SPICE validation using Cadence Spectre, where simulated performance is compared against specifications. If any specification is not met, COmPOSER tightens the corresponding constraint and re-initiates the synthesis flow from Stage~1, incorporating the updated margin.
This full re-synthesis approach ensures consistent convergence between schematic intent and post-layout behavior across performance metrics, without relying on manual tuning.
\vspace{-4mm}
\section{Experimental Setup and Results}
\label{sec:exp_results}
\noindent
We evaluate the use of COmPOSER to build CMOS RF/mm-wave LNAs and PAs in a commercial 65~nm technology, validated using Cadence EMX, Calibre extraction, and Spectre. We analyze the performance of the synthesized circuits--sizing, matching, linearity, stability--across frequency and COmPOSER runtime, and run an ablation study for the impact of the ML-based EM model versus layout-oblivious optimization. The COmPOSER open-source repository is available at \textit{\url{https://github.com/UMN-EDA/COmPOSER}}.

\noindent
\textbf{LNA Synthesis.}
COmPOSER is evaluated on four low noise amplifier designs: three single-stage LNAs at 
multiple frequencies (LNA~1--3) and a two-stage LNA (LNA~4) at the frequency of LNA-3.
The user specifies targets for $G$, $BW$, $S_{11}$, and $NF$, annotates RF-critical nets 
and constrains I/O pads to the north and west edges for transceiver-level integration. From these inputs, COmPOSER determines all active and passive dimensions in a single pass, requiring no constraint tuning or post-layout resizing. These sizes are then passed to the layout synthesizer, which generates the layouts. Fig.~\ref{fig:layouts}(a)–(c) show the layouts for three LNAs, with areas listed in an inset. The LNA-3 layout is omitted due to space limitations.

For each layout, Fig.~\ref{fig:smith_chart}(a) shows the input matching trajectories on a Smith chart, 
along the path A $\rightarrow$ B $\rightarrow$ C. Point A corresponds to $Z_{in,MOS}$, 
point B corresponds to the impedance after adding $L_g$,
and point C to the final matched impedance after insertion of the synthesized CPW (as a result of sweep in lines~\ref{step:lna_cpw_sweep}--\ref{step:lna_cpw_map} of Algorithm~\ref{algo:amp}).
Across all LNAs, point C lies within the $0.316$-radius circle corresponding to a 10~dB return-loss target, ensuring that over 90\%~\cite{razavi} of the incident signal is delivered. 

Post-layout verification confirms that the synthesized LNAs satisfy linearity requirements with high IIP3, and are shown to be stable even after PDN parasitics are included
($\kappa>1$). 
Table~\ref{tab:lna_performance_summary} shows post-layout EM- and SPICE-verified performance, area, and runtime, comparing against manual design where possible. We achieve comparable performance to manual design and reduce turnaround time from 70~hours to $<$24~minutes. Synthesis is very fast; $\sim$90\% of runtime is spent in final post-synthesis EM and SPICE verification.

{
\setlength{\abovecaptionskip}{-1pt}
\setlength{\belowcaptionskip}{-1pt}
\begin{figure}[t]
    \centering
    \vspace{-2mm}
    \includegraphics[width=0.85\linewidth]{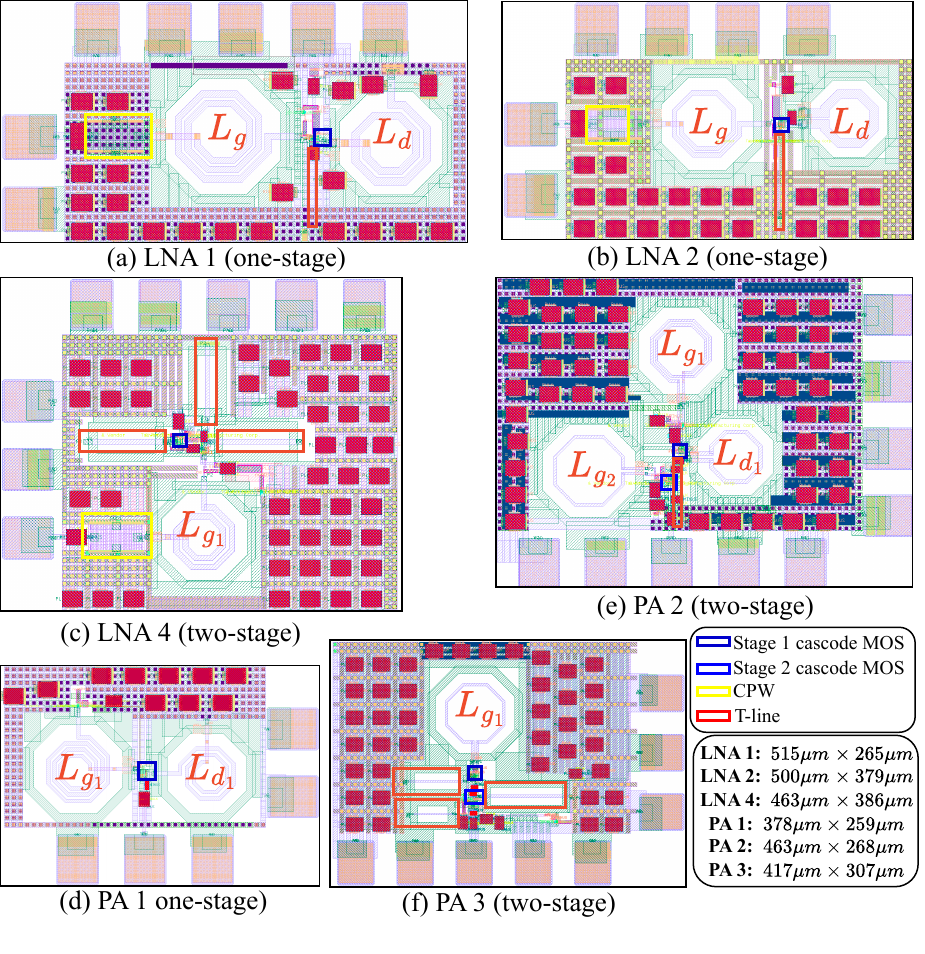}
    \vspace{-2mm}
    \caption{Layouts of LNA\{1,2,4\} and PA\{1,2,3\} (drawn to scale)}
    \Description{Layouts of LNAs and PAs (drawn to scale) \blueHL{You point out some sub-blocks but they are hard to see (except yellow) because the layout uses red and blue for other structures too.}}
    \label{fig:layouts}
\end{figure}
}

\newcolumntype{C}[1]{>{\centering\arraybackslash}m{#1}}
\begin{table}[t]
\vspace{-3mm}
\centering
\caption{Final post-layout performance of the LNA designs.}
\label{tab:lna_performance_summary}

\renewcommand{\arraystretch}{0.9}
\setlength{\tabcolsep}{2pt}

\begingroup
\resizebox{0.85\columnwidth}{!}{%
\begin{tabular}{|C{2.2cm}|C{1.35cm}|C{1.3cm}|C{1.3cm}|C{1.3cm}|C{1.2cm}|C{1.2cm}|}
\hline
\multirow{2}{*}{\textbf{Metric}}
& \multicolumn{2}{c|}{\textbf{LNA 1}}
& \multicolumn{2}{c|}{\textbf{LNA 2}}
& \textbf{LNA 3}
& \textbf{LNA 4} \\
\cline{2-7}
& \textbf{\cite{lna20ref}}
& \textbf{Our}
& \textbf{\cite{pa_lna40ref}}
& \textbf{Our}
& \textbf{Our}
& \textbf{Our} \\
\hline

CMOS process  & 65nm & 65nm & 65nm & 65nm & 65nm & 65nm \\ \hline
$f_{0}$ (GHz) & 22 & 25 & 39 & 45 & 60 & 60 \\ \hline
V$_{DD}$ (V)  & N/A & 1.1 & 1.1 & 1.1 & 1.1 & 1.1 \\ \hline

\multicolumn{1}{|C{2.2cm}|}{Topology}
& 1-stage cascode ($BW$ ext.)$^{*}$
& 1-stage cascode
& 1-stage cascode
& 1-stage cascode
& 1-stage cascode
& 2-stage cascode \\ \hline

$G$ (dB)  & 10.2 & 13.5 & 2.6 & 9.2 & 6.45 & 9.2 \\ \hline
$NF$ (dB) & 3.3  & 2.46 & 5.6 & 3.62 & 3.96 & 5.3 \\ \hline
$BW$ (GHz) & 14.5$^{*}$ & 6.1 & N/A & 4.8 & 4.7 & 4 \\ \hline

$S_{11}$ (dB) & -15 & -10 & -20 & -22 & -20 & -20 \\ \hline
$P_{dc}$ (mW) & 12.4 & 6.57 & 19.8 & 4.70 & 4.00 & 7.35 \\ \hline
IIP3 (dBm)    & -0.5 & -1.8 & N/A & -0.7 & 1.5 & -4 \\ \hline
Stable        & Yes  & Yes  & Yes & Yes & Yes & Yes \\ \hline
Chip area (mm$^2$) & 0.18 & 0.137 & N/A & 0.115 & 0.112 & 0.178 \\ \hline
Runtime (min) & N/A & 2.41 & N/A & 2.43 & 2.45 & 2.73 \\ \hline

\end{tabular}}
\endgroup

\vspace{3pt}
\raggedright
\footnotesize
\footnotesize{
$^{*}$\,\cite{lna20ref} uses $BW$-extension circuit, sacrificing $G$ and $NF$; we use standard cascode 
}

\end{table}

\begin{figure}[b]
    \vspace{2mm}
    \centering
    \setlength{\tabcolsep}{2pt} 
    \renewcommand{\arraystretch}{0}
    \subfloat[]{%
        \includegraphics[width=0.47\linewidth]{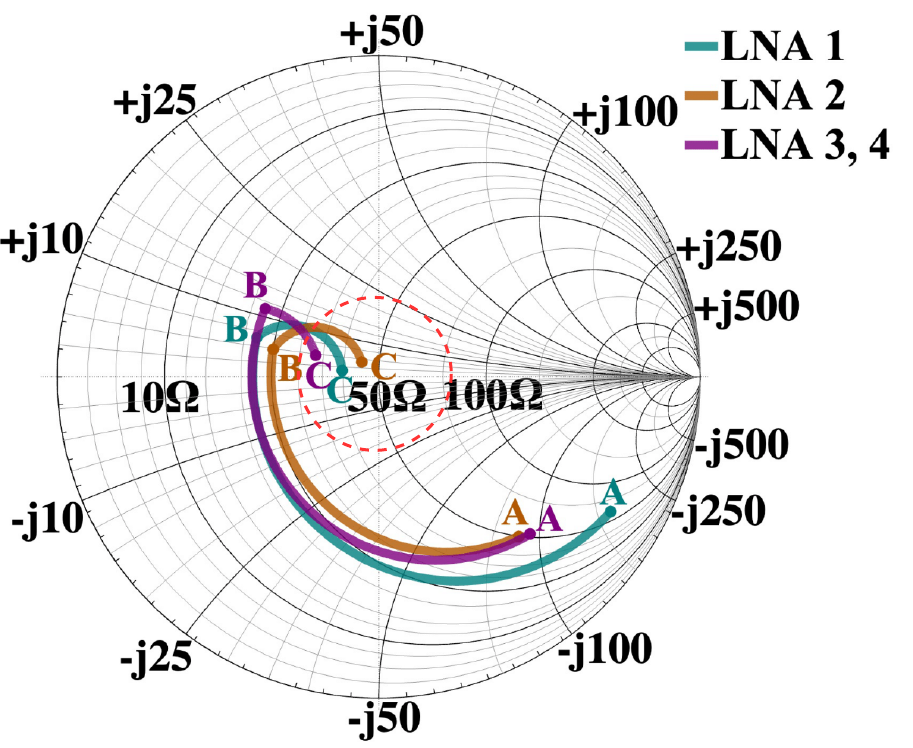}%
        \label{fig:lna_smith}
    }\hspace{-2mm} 
    \subfloat[]{%
        \includegraphics[width=0.41\linewidth]{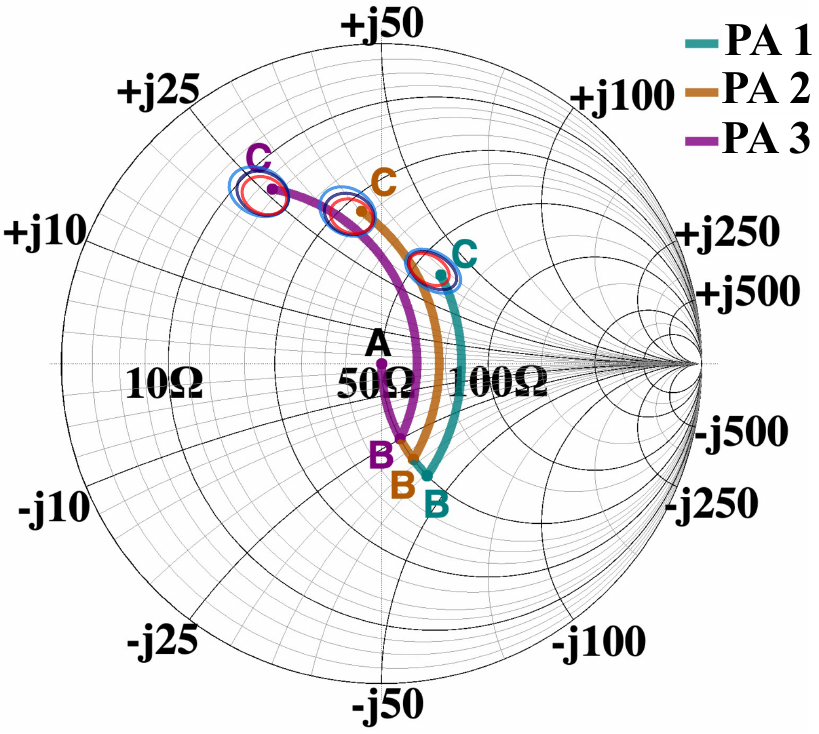}%
        \label{fig:pa_smith}
    }
    \vspace{-3mm}
    \caption{Smith chart showing (a) input matching of LNA designs [good design = within dotted orange circle] and (b)~output matching of PA designs [good design = within smallest oval]. In all cases, the final design C meets these criteria.}
    \Description{Smith chart showing (a) input matching of LNA designs and (b) output matching of PA designs.}
    \label{fig:smith_chart}
\end{figure}

\noindent
\textbf{PA Synthesis.}
Next, we focus on Class~A power amplifiers: PA-1 (single stage) at 20~GHz, and PA-2 and PA-3 (both two-stage) at 40 and 60~GHz, respectively. COmPOSER determines the number of PA stages based on the target $P_{sat}$, $G$, and load-line conditions (line~\ref{step:PA_stage_scale}).
COmPOSER generates dimensions for all actives and passives and synthesizes the layouts shown in Fig.~\ref{fig:layouts}(d)--(f).

Building on these layouts, Fig.~\ref{fig:smith_chart}(b) shows the output load transformation trajectories. The paths move from the 50~$\Omega$ pad impedance at A (this initial point is identical for all PA designs), through $C_{out}$ shift to B, and then to C via shunt-inductor ($L_{d1}$ for PA-1 or $L_{d2}$ for PA-2,3) that completes the match (line~\ref{step:pa_match}).
Point~C is the synthesized load seen by the final drain. The colored ovals near Point~C are output power contours: each is the locus of impedances giving a fixed output power, with higher-power contours being smaller than lower-power contours~\cite{sorin_book, razavi}. 
COmPOSER 
achieves a final design point C inside the tight high-power contours, achieving high $P_{sat}$.

All synthesized PAs satisfy their specification targets after post-layout evaluation. Even after including PDN parasitics, the designs are stable, with Rollett stability factor $\kappa>1$.
Table~\ref{tab:pa_multi_freq_summary} reports the post-layout SPICE and EM-validated performance along with area and runtime. COmPOSER reduces the total turnaround from several hours to roughly three minutes for synthesis, and forty minutes in all, including performance verification.

\newcolumntype{C}[1]{>{\centering\arraybackslash}m{#1}}
\begin{table}[t]
\centering
\caption{Final post-layout performance of the PA designs.}
\label{tab:pa_multi_freq_summary}
\renewcommand{\arraystretch}{0.85}
\setlength{\tabcolsep}{1pt}
\begingroup
\resizebox{!}{1.8cm}{
\begin{tabular}{|C{2.5cm}|C{1.9cm}|C{1.9cm}|C{1.9cm}|}
    \hline
    \textbf{Metric}
    & \textbf{PA 1} 
    & \textbf{PA 2} 
    & \textbf{PA 3} \\
    \hline

    CMOS process & 65nm & 65nm & 65nm \\ \hline
    $f_0$ (GHz)  & 20 & 40 & 60 \\ \hline
    V$_{DD}$ (V) & 1.2 & 1.2 & 1.2 \\ \hline

    Topology
    & 1-stage cascode
    & 2-stage cascode
    & 2-stage cascode \\ \hline

    $G$ (dB)            & 14.7 & 16 & 13.6 \\ \hline
    $P_{\text{SAT}}$ (dBm) & 9.5 & 6.7 & 6.8 \\ \hline
    $BW$ (GHz)          & 10.2 & 5.1 & 8.3 \\ \hline
    $P_{dc}$ (mW)       & 51.3 & 53.0 & 63.5 \\ \hline
    PAE$_{max}$ (\%)    & 25 & 7.5 & 6.4 \\ \hline
    Stable              & Yes & Yes & Yes \\ \hline
    Chip area (mm$^2$)  & 0.220 & 0.254 & 0.282 \\ \hline
    Runtime (min)       & 2.31 & 3.25 & 3.74 \\ \hline
\end{tabular}}
\endgroup
\end{table}

\begin{table}[t]
\centering
\caption{Average COmPOSER runtime vs. manual LNA design.}
\label{tab:runtime_breakdown}
\renewcommand{\arraystretch}{0.9}
\setlength{\tabcolsep}{1pt}
\scriptsize
\resizebox{\linewidth}{!}{
\begin{tabular}{|>{\centering\arraybackslash}m{3cm}|c|c|>{\centering\arraybackslash}m{2cm}|}
\hline
\textbf{Task} & \textbf{LNA} & \textbf{PA} & \textbf{Manual LNA} \\
\hline

\multirow{2}{*}{
\parbox{3cm}{\centering
\textbf{One-time} ML-based EM model
(for Inductor, T-line, CPW)}}
& \multicolumn{2}{c|}{$\approx$24 hrs (data gen.)} & \multirow{3}{*}{\parbox{2cm}{\centering Schematic optimization\\5~hrs }}\\
& \multicolumn{2}{c|}{$<1$ min (training)} & \\
\cline{1-3}

Stage 1 -- Hybrid optimization & 52 sec & 38 sec & \\
\hline

Stage 2 -- Primitive generation & 35 sec & 61 sec &\multirow{2}{*}{\parbox{2cm}{\centering Layout generation\\ 25~hrs}} \\

\cline{1-3}

Stage 3 -- Layout synthesis & 68 sec & 87 sec &  \\
\hline

Stage 4 -- SPICE and EM simulation & 21 min & 37 min & \multirow{3}{*}{\parbox{2cm}{\centering Post-layout optimization\\ 40~hrs}} \\
\cline{1-3}

\multirow{2}{*}{
\parbox{3cm}{\centering
Subtotal (Stages 1--3 only)\\
\textit{(design generation time only)}}}
& \textbf{2.58 min} & \textbf{3.10 min} & \\
& 155 sec & 186 sec & \\
\hline

\textbf{Total runtime} (COmPOSER + EM/SPICE)
& \textbf{23.58 min} & \textbf{40.10 min} & \textbf{$\sim$70 hrs} \\
\hline

\textbf{Runtime improvement (LNA)}
& \multicolumn{2}{c|}{} & \textbf{$\sim$178$\times$ faster }\\
\hline

\end{tabular}}
\end{table}

\noindent
\textbf{EM ablation study and runtime.}
To demonstrate the risks of layout-oblivious sizing, we replace Stage 1 with a schematic-only flow. 
The intended 60~GHz center frequency is severely degraded to 45~GHz for the PA and 40~GHz for the LNA due to unmodeled parasitics. Performance at the target band deteriorates sharply: the PA suffers 4.6~dB collapse in $G$ and a 4~dB drop in output power; the LNA sees 3.4~dB reduction in $G$ alongside a 1.75~dB $NF$ degradation.

Restoring EM-aware modeling 
achieves exact target alignment for the PAs and the 25~GHz LNA, with only minor systematic offsets in LNA-3. Stages~1–3 take minutes (Table~\ref{tab:runtime_breakdown}), and the final full-wave EM and SPICE verification (Stage~4) dominates the runtime. Using a manual 40~GHz LNA design ($\approx$70~hours) as a baseline, COmPOSER's generation time ($\approx$25~minutes, $>80\%$ of which is required for final performance verification) provides a 100--300$\times$ productivity gain.

\vspace{-2mm}
\section{Conclusion}
\label{sec:conclusion}

\noindent
COmPOSER synthesizes layout-aware optimized implementations of LNAs and PAs at multiple RF/mm-wave frequencies based on user specifications. The approach employs accurate ML-based design of passives, which form the major performance bottleneck, and builds a flow that combines the optimization of active devices, passives, and matching networks, and provides a placed-and-routed solution, including routing to pads and power grid routing. The solution quality is demonstrated through performance comparisons with manual design and Smith chart trajectories, and the results show orders of magnitude speedup over manual design at iso-quality.

\balance
\bibliographystyle{misc/IEEEtran}
\bibliography{bib/main.bib}

@string{cad = "IEEE T. Comput. Aid D."}

@string{jssc = "IEEE J. Solid-St. Circ."}

@string{iccad = "Proc. ICCAD"}

@string{rfic = "Proceedings of the IEEE Radio Frequency Integrated Circuits Symposium"}

@string{itit  = "IEEE Transactions on Information Theory"}

@string{dac = "Proc. DAC"}

@string{iojcas = "IEEE Open Journal of Circuits and Systems"}

@string{date = "Proc. DATE "}

@string{rfit = "Proceedings of the IEEE International Symposium on Radio-Frequency Integration Technology"}

@string{tmtt ="IEEE Transactions on Microwave Theory and Techniques"}

@string{cas2 = "IEEE T. Circuits Syst. II"}

@string{cas1 = "IEEE T. Circuits Syst. I"}

@string{mwcl = "IEEE Microwave and Wireless Components Letters"}

@string{tssc = "IEEE Transactions on Systems Science and Cybernetics"}

@string{dnt = "IEEE Des. Test"}

@string{iccd = "Proc. ICCD"}

@string{mttsd = "IEEE MTT-S International Microwave Symposium Digest"}

@string{mtts="Proceedings of the IEEE/MTT-S International Microwave Symposium"}

@string{cad = "IEEE Transactions on Computer-Aided Design of Integrated Circuits and Systems"}

@string{cas1 = "IEEE Transactions on Circuits and Systems I"}

@string{cas2 = "IEEE Transactions on Circuits and Systems II"}

@string{jssc = "IEEE Journal of Solid-State Circuits"}

@string{iccad = "Proceedings of the IEEE/ACM International Conference on Computer-Aided Design"}

@string{dac = "Proceedings of the ACM/IEEE Design Automation Conference"}

@string{date = "Proceedings of the Design, Automation \& Test in Europe Conference"}

@string{iccd = "Proceedings of the IEEE International Conference on Computer Design"}

@string{dnt = "IEEE Design \& Test"}

@string{cambupress = "Cambridge University Press"}

@string{iretct = "IRE Transactions on Circuit Theory"}

@ARTICLE{sorin_2007,
  author={Yao, Terry and Gordon, Michael Q. and Tang, Keith K. W. and Yau, Kenneth H. K. and Yang, Ming-Ta and Schvan, Peter and Voinigescu, Sorin P.},
  journal=jssc, 
  title={Algorithmic Design of {CMOS LNAs} and {PAs} for 60-{GHz} Radio}, 
  year=2007,
  month = may,
  pages="1044--1057",
  volume={42},
  number={5}}

@INPROCEEDINGS{mansour_2005,
  author={Miraftab, V. and Mansour, R.R.},
  booktitle=mttsd, 
  title={Tuning of microwave filters by extracting human experience using fuzzy logic}, 
  year=2005,
  volume={},
  number={},
  pages="1605--1608"}

@INPROCEEDINGS{crols_95,
  author={Crols, J. and Donnay, S. and Steyaert, M. and Gielen, G.},
  booktitle=iccad, 
  title={A high-level design and optimization tool for analog {RF} receiver front-ends}, 
  year=1995,
  volume={},
  number={},
  pages={550-553},
  doi={10.1109/ICCAD.1995.480170}}

@inproceedings{gielen_2000,
author = {Vancorenland, Peter and De Ranter, C. and Steyaert, M. and Gielen, G.},
title = {Optimal {RF} design using smart evolutionary algorithms},
year = 2000,
booktitle = dac,
pages = "7--10",
numpages = {4}
}

@ARTICLE{sansen_2002,
  author={De Ranter, C.R.C. and Van der Plas, G. and Steyaert, M.S.J. and Gielen, G.G.E. and Sansen, W.M.C.},
  journal=cad, 
  title={{CYCLONE:} {A}utomated design and layout of {RF LC}-oscillators}, 
  year=2002,
  volume={21},
  number={10},
  month = Oct,
  pages="1161--1170"}

@ARTICLE{taiyun_2023,
  author={Hu, Yaolong and Chi, Taiyun},
  journal=cas1, 
  title={A Systematic Approach to Designing Broadband Millimeter-Wave Cascode Common-Source With Inductive Degeneration Low Noise Amplifiers}, 
  year=2023, 
  month = Jan,
  volume={70},
  number={4},
  pages="1489--1502"}

@ARTICLE{pileggi_2007,
  author={Li, Xin and Gopalakrishnan, Padmini and Xu, Yang and Pileggi, Lawrence T.},
  journal=cad, 
  title={Robust Analog/{RF} Circuit Design With Projection-Based Performance Modeling}, 
  month = Jan,
  year=2007,
  volume={26},
  number={1},
  pages="2--15"}

@ARTICLE{gielen_2014,
  author={Liu, Bo and Zhao, Dixian and Reynaert, Patrick and Gielen, Georges G. E.},
  journal=cad, 
  title={{GASPAD}: A General and Efficient mm-Wave Integrated Circuit Synthesis Method Based on Surrogate Model Assisted Evolutionary Algorithm}, 
  year=2014,
  volume={33},
  number={2},
  pages="169--182",
  month = Jan}

@INPROCEEDINGS{zeng_2018,
  author={Lyu, Wenlong and Yang, Fan and Yan, Changhao and Zhou, Dian and Zeng, Xuan},
  booktitle=dac, 
  title={Multi-objective {Bayesian} Optimization for Analog/{RF} Circuit Synthesis}, 
  year=2018,
  volume={},
  number={},
  note="(6 pages)"}

@INPROCEEDINGS{ricardo_2018,
  author={Pessoa, Tiago and Lourenço, Nuno and Martins, Ricardo and Póvoa, Ricardo and Horta, Nuno},
  booktitle=date, 
  title={Enhanced analog and {RF IC} sizing methodology using {PCA} and {NSGA-II} optimization kernel}, 
  year=2018,
  pages="660--665"}

@ARTICLE{zhang_2022,
  author={Feng, Feng and Na, Weicong and Jin, Jing and Zhang, Jianan and Zhang, Wei and Zhang, Qi-Jun},
  journal=tmtt, 
  title={Artificial Neural Networks for Microwave Computer-Aided Design: The State of the Art}, 
  year=2022,
  month = Aug,
  volume={70},
  number={11},
  pages="4597--4619"}

@ARTICLE{zeng_2022,
  author={Huang, Jiangli and Tao, Cong and Yang, Fan and Yan, Changhao and Zhou, Dian and Zeng, Xuan},
  journal=tmtt, 
  title={Bayesian Optimization Approach for {RF} Circuit Synthesis via Multitask Neural Network Enhanced {Gaussian} Process}, 
  year=2022,
  month= Aug,
  volume={70},
  number={11},
  pages="4787--4795"}

@inproceedings{ma_2022,
  author = {Cao, Weidong and Benosman, Mouhacine and Zhang, Xuan and Ma, Rui},
  booktitle = dac,
  title = {Domain knowledge-infused deep learning for automated analog/radio-frequency circuit parameter optimization},
  year = 2022,
  pages = "1015-–1020",
  numpages = {6}}

@INPROCEEDINGS{wang_2021,
  author={Er, Siawpeng and Liu, Edward and Chen, Minshuo and Li, Yan and Liu, Yuqi and Zhao, Tuo and Wang, Hua},
  booktitle=mtts, 
  title={Deep Learning Assisted End-to-End Synthesis of mm-Wave Passive Networks with {3D EM} Structures: {A} Study on A Transformer-Based Matching Network}, 
  year=2021,
  volume={},
  number={},
  pages={66--69}}

@inproceedings{pan_2025,
  author = {Chae, Hyunsu and Yu, Hao and Li, Sensen and Pan, David Z.},
  title = {{PulseRF}: Physics Augmented {ML} Modeling and Synthesis for High-Frequency {RFIC} Design},
  year = 2024,
  booktitle  = iccad,
  note="(9 pages)"}

@INPROCEEDINGS{wang_2020,
  author={Munzer, David and Er, Siawpeng and Chen, Minshuo and Li, Yan and Mannem, Naga S. and Zhao, Tuo and Wang, Hua},
  booktitle=rfic, 
  title={Residual Network Based Direct Synthesis of {EM} Structures: A Study on One-to-One Transformers}, 
  year=2020,
  volume={},
  number={},
  pages="143--146"}

@inproceedings{chi_2025,
  author = {Hu, Yaolong and Guo, Hao and Wang, Shikai and Liu, Jiaqi and Cao, Weidong and Chi, Taiyun},
  year = 2025,
  note="(7 pages)",
  title = {{AdreamDCO:} {AI}-Driven Robust and Efficient Design Automation for Digitally Controlled Oscillators}, 
  booktitle = dac}

@misc{gurobi,
  author = {{Gurobi Optimization, LLC}},
  title = {{Gurobi Optimizer Reference Manual}},
  year = 2024,
  url = "https://www.gurobi.com"
}

@ARTICLE{mendes_2025,
  author={Mendes, Luís and Silva, João and Lourenço, Nuno and Vaz, João Caldinhas and Martins, Ricardo and Passos, Fábio},
  journal=tmtt, 
  title={Fully Automatically Synthesized mm-Wave Low-Noise Amplifiers for {5G/6G} Applications}, 
  year=2025,
  pages="4828--4841"}

@INPROCEEDINGS{ghosh_2025,
  author={Ghosh, Subhadip and Gebru, Endalk Y. and Kashyap, Chandramouli V. and Harjani, Ramesh and Sapatnekar, Sachin S.},
  booktitle=date, 
  title={Accelerating {OTA} Circuit Design: Transistor Sizing Based on a Transformer Model and Precomputed Lookup Tables}, 
  year=2025,
  volume={},
  number={},
  note="(7 pages)"}

@inproceedings{kunal_2019,
author = {Kunal, Kishor and Madhusudan, Meghna and Sharma, Arvind K. and Xu, Wenbin and Burns, Steven M. and Harjani, Ramesh and Hu, Jiang and Kirkpatrick, Desmond A. and Sapatnekar, Sachin S.},
title = {{ALIGN:} {O}pen-Source Analog Layout Automation from the Ground Up},
year = 2019,
booktitle = dac,
note="(4 pages)"
}

@inproceedings{Wang_2020_gcnrl,
   title={{GCN-RL} Circuit Designer: Transferable Transistor Sizing with Graph Neural Networks and Reinforcement Learning},
   booktitle=dac,
   author={Wang, Hanrui and Wang, Kuan and Yang, Jiacheng and Shen, Linxiao and Sun, Nan and Lee, Hae-Seung and Han, Song},
   year={2020},
   note="(6 pages)"
}

@INPROCEEDINGS{Abel_19,
  author={Abel, Inga and Neuner, Maximilian and Graeb, Helmut},
  booktitle=iccd, 
  title={Constraint-Programmed Initial Sizing of Analog Operational Amplifiers}, 
  year={2019},
  pages={413-421},
  doi={10.1109/ICCD46524.2019.00065}}

@inproceedings{settaluri_20,
author = {Settaluri, Keertana and Haj-Ali, Ameer and Huang, Qijing and Hakhamaneshi, Kourosh and Nikolic, Borivoje},
title = {{AutoCkt}: Deep reinforcement learning of analog circuit designs},
year = {2020},
booktitle = date,
pages = "490-495",
}

@ARTICLE{hao_2021,
  author={Chen, Hao and Liu, Mingjie and Xu, Biying and Zhu, Keren and Tang, Xiyuan and Li, Shaolan and Lin, Yibo and Sun, Nan and Pan, David Z.},
  journal=dnt, 
  title={{MAGICAL:} {A}n Open- Source Fully Automated Analog {IC} Layout System from Netlist to {GDSII}}, 
  year=2021,
  pages="19--26"}

@ARTICLE{guo_2025,
  author={Guo, Felicia and Zhou, Bob and Biswas, Ayan and Kwon, Paul and Liu, Zhaokai and Ho, Ken and Stojanović, Vladimir and Nikolić, Borivoje},
  journal=iojcas, 
  title={{BAG3++:} {A}n Extensible Generator Framework for Automated Layout-Aware {AMS} Design}, 
  year=2025,
  pages="181--191"}

@book{sorin_book,
  title={High-Frequency Integrated Circuits},
  author={Voinigescu, S.},
  year=2013,
  publisher=cambupress,
  address="Cambridge, UK"
}

@book{wolsey_integer_programming_2020,
  author    = {Laurence A. Wolsey},
  title     = {Integer Programming},
  year      = 2020,
  publisher = {John Wiley \& Sons, Inc.},
  address   = "New Jersey, USA"
}

@article{hart_astar_1968,
  author    = {Peter E. Hart and Nils J. Nilsson and Bertram Raphael},
  title     = {A Formal Basis for the Heuristic Determination of Minimum Cost Paths},
  journal   = tssc,
  volume    = {SSC-4},
  number    = {2},
  pages     = "100--107",
  year      = 1968
}

@book{razavi,
author = {Razavi, Behzad},
title = {{RF} Microelectronics},
year = 2011,
edition = "2nd",
publisher = {Prentice Hall Press},
address = "New Jersey, USA",
}

@book{Lee_2003, 
address="Cambridge, UK", 
title={The Design of {CMOS} Radio-Frequency Integrated Circuits}, 
edition = {2nd},
publisher=cambupress, 
author={Lee, Thomas H.}, 
year=2003
}

@ARTICLE{cpwref,
  author={Cheung, T.S.D. and Long, J.R.},
  journal=jssc, 
  title={Shielded passive devices for silicon-based monolithic microwave and millimeter-wave integrated circuits}, 
  year=2006,
  volume={41},
  number={5},
  pages="1183-1200"}

@ARTICLE{chen_2012,
  author={Niknejad, Ali M. and Chowdhury, Debopriyo and Chen, Jiashu},
  journal=tmtt, 
  title={Design of {CMOS} Power Amplifiers}, 
  year=2012,
  volume={60},
  number={6},
  pages="1784-1796"}

@ARTICLE{lee_2002,
  author={Jung-Suk Goo and Hee-Tae Ahn and Ladwig, D.J. and Zhiping Yu and Lee, T.H. and Dutton, R.W.},
  journal=jssc, 
  title={A noise optimization technique for integrated low-noise amplifiers}, 
  year="2002",
  volume={37},
  number={8},
  pages="994-1002"}

@ARTICLE{sengupta_2023,
  author={Karahan, Emir Ali and Liu, Zheng and Sengupta, Kaushik},
  journal=jssc, 
  title={Deep-Learning-Based Inverse-Designed Millimeter-Wave Passives and Power Amplifiers}, 
  year=2023,
  volume={58},
  number={11},
  pages="3074-3088"}

@ARTICLE{niknejad_2008,
  author={Chowdhury, Debopriyo and Reynaert, Patrick and Niknejad, Ali M.},
  journal=cas2, 
  title={Transformer-Coupled Power Amplifier Stability and Power Back-Off Analysis}, 
  year=2008,
  volume={55},
  number={6},
  pages="507-511"}

@ARTICLE{rollett_1962,
  author={Rollett, J.},
  journal=iretct, 
  title={Stability and Power-Gain Invariants of Linear Twoports}, 
  year=1962,
  volume={9},
  number={1},
  pages="29-32"}

@INPROCEEDINGS{pa_lna40ref,
  author={Zhang, Xuexue and Qiao, Kun and Chen, Qin and Liang, Yue and Li, Lianming and Feng, Jun},
  booktitle=rfit, 
  title={A 39 {GHz T/R} Front-End Module in 65nm {CMOS}}, 
  year=2021,
  volume={},
  number={},
  note="(3 pages)"}

@ARTICLE{lna20ref,
  author={Qin, Pei and Xue, Quan},
  journal=mwcl, 
  title={Compact Wideband {LNA} With Gain and Input Matching Bandwidth Extensions by Transformer}, 
  year=2017,
  volume={27},
  number={7},
  pages="657-659"}

@article{random_forest,
author = {Breiman, Leo},
title = {Random Forests},
year = {2001},
issue_date = {October 1 2001},
publisher = {Kluwer Academic Publishers},
address = {USA},
volume = {45},
number = {1},
journal = {Machine Learning},
month = oct,
pages = "5–32"
}

@ARTICLE{knn,
  author={Cover, T. and Hart, P.},
  journal=itit, 
  title={Nearest neighbor pattern classification}, 
  year=1967,
  volume={13},
  number={1},
  pages={21-27}}

@INPROCEEDINGS{hpwl,
  author={C.-w. Sham and Young, E.R.Y. and Chu, C.},
  booktitle=dac, 
  title={Optimal cell flipping in placement and floorplanning}, 
  year=2006,
  pages="1109-1114"}

@ARTICLE{Kim03,
  author={Jae-Gon Kim and Yeong-Dae Kim},
  journal=cad, 
  title={A linear programming-based algorithm for floorplanning in {VLSI} design}, 
  year={2003},
  volume={22},
  number={5},
  pages={584-592},
}

@ARTICLE{staircase_MOS,
  author={Bücher, Thomas and Grzyb, Janusz and Hillger, Philipp and Rücker, Holger and Heinemann, Bernd and Pfeiffer, Ullrich R.},
  journal=jssc, 
  title={A Broadband 300 {GHz} Power Amplifier in a 130 nm {SiGe BiCMOS} Technology for Communication Applications}, 
  year=2022,
  volume={57},
  number={7},
  pages="2024-2034"}

\end{document}